\documentclass[oupdraft]{bio}
\usepackage{url}
\usepackage{xcolor}
\usepackage{float}

\begin{document}

\title{Estimating menarcheal age distribution from partially recalled data}
\author{Sedigheh Mirzaei Salehabadi$^\ast$, Debasis Sengupta, Rahul Ghosal\\[4pt]
\textit{St. Jude Children's Research Hospital, Memphis, USA\\
Indian Statistical Institute, Kolkata, India\\
North Carolina State University, Raleigh, USA}\\
[2pt]
{Sedigheh.Mirzaei@stjude.org}
}

\markboth%
{S. Mirzaei S. and others}
{Estimating menarcheal age distribution from partially recalled data}

\maketitle


\begin{abstract}
{In a cross-sectional study, adolescent and young adult females were asked to recall the time of menarche, if experienced. Some respondents recalled the date exactly, some recalled only the month or the year of the event, and some were unable to recall anything. We consider estimation of the menarcheal age distribution from this interval censored data. A~complicated interplay between age-at-event and calendar time, together with the evident fact of memory fading with time, makes the censoring informative. We propose a model where the probabilities of various types of recall would depend on the time since menarche. For parametric estimation we model these probabilities using multinomial regression function. Establishing consistency and asymptotic normality of the parametric MLE requires a bit of tweaking of the standard asymptotic theory, as the data format varies from case to case. We also provide a non-parametric MLE, propose a computationally simpler approximation, and establish the consistency of both these estimators under mild conditions. We study the small sample performance of the parametric and non-parametric estimators through Monte Carlo simulations.  Moreover, we provide a graphical check of the assumption of the multinomial model for the recall probabilities, which appears to hold for the menarcheal data set. Our analysis shows that the use of the partially recalled part of the data indeed leads to smaller confidence intervals of the survival function.}
{Interval censoring, Informative censoring,
Maximum likelihood estimation, Retrospective study, Current status data, Self consistency.}
\end{abstract}

\section{Introduction}
\label{intro}
In a recent survey conducted by the Indian Statistical Institute (ISI) in and around the city of Kolkata \citep{Dasgupta_2015}, over four thousand randomly selected individuals, aged between 7 and 21 years, were sampled. In this retrospective and cross-sectional study, the subjects were interviewed on or around their birthdays. The data on female subjects contains age, menarcheal status, some physical measurements and information on some socioeconomic variables. If a subject had already experienced menarche, she was asked to recall the date of the onset of her menarche.


We considered a subset of the original data, consisting of respondents who came from a general caste family with monthly expenditure greater than or equal to Rs. 15000 and both parents graduate.
Among the 289 females represented in the data set, 45 individuals did not have menarche, 68 individuals recalled the exact date of the onset of menarche, 43 and 30 individuals recalled the calendar month and the calendar year of the onset, respectively, and 103 individuals could not recall any range of dates.
Thus, the data are interval-censored. A major goal of this study is to estimate the distribution of the age at onset of menarche.

This problem should be of interest to anyone working with incompletely recalled time-to-event data, of which there are many examples in the literature. The key variables in these studies include age at onset of menarche in adolescent and young adult females \citep{Koo_1997}, time-to-pregnancy \citep{Joffe_1995}, time-to-weaning from breastfeeding \citep{Gillespie_2006}, time-to-injury for victims injured during a year \citep{Harel_1994}, time-to-employment \citep{Mathiowetza_1988}, and so on.
In these studies, estimation of the time-to-event distribution is important for building a standard for individuals, comparing two populations or assessing the effect of a covariate. There is a possibility that the recalled time-to-event is inaccurate \citep{Koo_1997,Mathiowetza_1988}. In the ISI study, this problem was somewhat circumvented by allowing the respondents to report an interval in lieu of the exact age-at-menarche. The recalled intervals generally happened to be in terms of calendar months and years. We refer to this special type of incompleteness as partial recall.

\begin{figure}[t]
\centering
 \includegraphics[height=2.8in,width=3.6in]{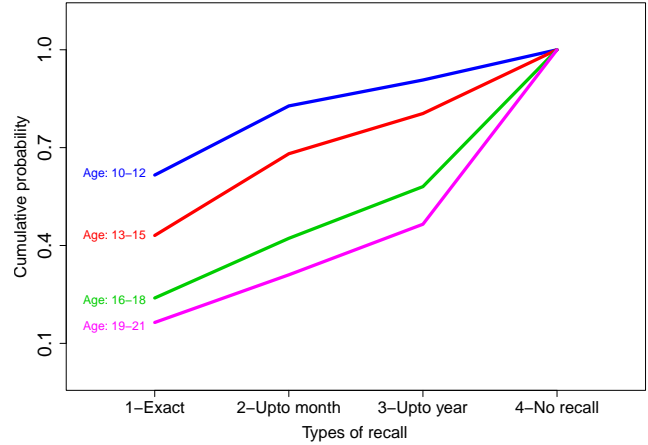}
\caption[]{Cumulative proportion of decreasing degrees of recall for different age ranges in menarcheal data} \label{fig_1}
\end{figure}

Figure~\ref{fig_1} shows the cumulative proportions of successively less precise recall in different groups of ages at interview, for the respondents of the ISI study. It is seen that the lines do not cross and the age group order is preserved. Also, there is greater precision of recall at lower age group, i.e., memory fades with time. Thus, two subjects interviewed at the same age would have different chances of recalling their age at menarche, depending on which of them had experienced the event earlier. In other words, the censoring mechanism underlying such recall-based data is inherently informative. The natural question is: how can one model the different degrees of partial recall, so that the distribution of menarcheal age can be estimated?

There is no suitable model and method in the literature for estimating the time-to-event distribution from partially recalled data, though such data abound in various fields. Apart from the informative nature of censoring, the problem is complicated by the mismatch of the time scales of the partial recall information (expressed through calendar time) and the time-to-event (measured from a respondent-specific starting time, e.g., birth). \cite{Mirzaei_2015} and \cite{Mirzaei_2016} addressed the first issue by proposing a model for this type of informative censoring, but they bypassed the second issue by clubbing all the cases of partial recall with the cases of no recall.

In this paper we propose a realistic censoring model for estimating the time-to-event distribution from partially recalled data.
We present our modelling framework in Section~\ref{likelihood}, and derive the appropriate likelihood under the proposed model. In Section~\ref{Param} we express the likelihood as a product of densities in an appropriate space, and discuss asymptotic properties of a parametric maximum likelihood estimator (MLE). In Section~\ref{nonparametric} we derive the non-parametric maximum likelihood estimator (NPMLE) and an approximate MLE (AMLE), and also establish consistency of both these estimators. In Sections~\ref{Simulation} and~\ref{adequacy} we report the results of Monte Carlo simulations of small sample performance of the MLE and the AMLE, and present some diagnostic checks of adequacy of the model. We analyze the real data set in Section~\ref{DataAnalysis}. We conclude with some discussion and indications of possible future extensions in Section \ref{s:discuss}. The proofs of all the theorems and the results of additional simulations and data analysis are given in the supplementary material.

\section{Model and Likelihood}
\label{likelihood}
Consider a set of $n$ subjects having ages at occurrence of landmark events $T_1,\ldots,T_n$, which are samples from the distribution $F$, with density $f$. Let these subjects be interviewed at ages $S_1,\ldots,S_n$, respectively.
Suppose the $S_i$s are samples from another distribution and are independent of the $T_i$s. Let $\delta_i$
be the indicator of $T_i\le S_i$.
This inequality means that the event for the $i$th subject had occurred
on or before the time of interview.

In the case of current status data, one only observes
$(S_i,\delta_i), i=1,2,\ldots, n$. The corresponding likelihood,
conditional on the times of interview, is
\begin{equation}
\prod_{i=1}^{n}[F(S_i)]^{\delta_i}[\bar{F}(S_i)]^{1-\delta_i},  \label{YN}
\end{equation}
where $\bar{F}(S_i)=1-F(S_i)$. For properties of the MLE based on the above likelihood, see \cite{Lee_2003}.

The structure of recalled data is generally more complicated. \cite{Mirzaei_2015} proposed a simplistic model, where the subject may either recall the time of the event exactly or not remember it at all. They used an indicator, $\varepsilon_i$, to record whether an exact recall is possible. As the chance of recall may depend on the time elapsed since the event, they modeled the non-recall probability as a function of this time. According to
this model,
\begin{equation*}
P(\varepsilon_i=0|S_i=s,T_i=t)=\pi(s-t) \qquad \mbox{for} \quad 0<t<s, \label{pifirst}
\end{equation*}
for some non-recall probability function $\pi$. Thus, the likelihood is
\begin{equation}
\prod_{i=1}^{n}\left[\left(\int_{0}^{S_i} f(u)\pi(S_i-u) du\right)^{1-\varepsilon_i}\left[f(T_i)(1-\pi(S_i-T_i))\right]^{\varepsilon_i}\right]^{\delta_i}
[\bar{F}(S_i)]^{1-\delta_i}. \label{MirSen}
\end{equation}

Let us now consider the possibility that the $i$th subject can recall the date of the event only up to a calendar month or a
calendar year, and define the recall status variable $\varepsilon_i$ for the $i$th subject as
\begin{equation}
\varepsilon_i= \left\{
\begin{array}{ll}
0 & \text {if there is exact recall},\\
1 & \text {if the date is recalled up to the calendar month},\\
2 & \text {if the date is recalled up to the calendar year},\\
3 & \text {if the date is not recalled}.
\end{array} \right. \label{epsilon}
 \end{equation}
The value of $\varepsilon_i$ concerns the state of recall. When $\delta_i=1$, $\varepsilon_i=0$ means that the exact date of the event is recalled. When $\delta_i=0$, $\varepsilon_i$ may be assigned the value 0, as no recall failure is expected in case the event is reported not to have happened.

We regard the four scenarios as outcomes of a multinomial selection, where allocation probabilities depend on the time elapsed since the occurrence of the event. Thus, for $0<t<s$, we model the allocation probabilities as
\begin{equation}
 \begin{array}{l}
P(\varepsilon_i=0|S_i=s,T_i=t)=\pi^{(0)}(s-t),\\
P(\varepsilon_i=1|S_i=s,T_i=t)=\pi^{(1)}(s-t),\\
P(\varepsilon_i=2|S_i=s,T_i=t)=\pi^{(2)}(s-t),\\
P(\varepsilon_i=3|S_i=s,T_i=t)=\pi^{(3)}(s-t).
\end{array} \label{pifunc}
\end{equation}
where $\sum_{k=0}^{3}\pi^{(k)}(s-t)=1$.

We refer to the set-up described in the first paragraph of this section, together with \eqref{epsilon} and \eqref{pifunc} as the proposed model. According to this model, contributions to the likelihood in different cases are as follows.
\begin{description}
 \item[{\sc Case}] (i) When $\delta_i=0$ (the event has not occurred till the time of observation), the contribution of the $i$th individual to the likelihood is~$\bar{F}(S_i)$.
\item[{\sc Case}] (ii): When $\delta_i=1$ and $\varepsilon_i=0$ (the event has occurred and the $i$th individual can remember the time), the contribution of the individual to the likelihood is $f(T_i)\pi^{(0)}(S_i-T_i)$.
\item[{\sc Case}] (iii): When $\delta_i=1$ and $\varepsilon_i=1$ (the event has occurred but the $i$th individual can only recall the  calendar month of the event), the contribution of the individual to the likelihood is $\int_{M_{i1}}^{M_{i2}} f(u)\pi^{(1)}(S_i-u) du$, where $M_{i1}$ and $M_{i2}$ are the ages of the individual at the beginning and the end of the calendar month recalled by the individual.
\item[{\sc Case}] (iv): When $\delta_i=1$ and $\varepsilon_i=2$ (the event has occurred but the $i$th individual can only recall the calendar year of the event), the contribution of the individual to the likelihood is $\int_{Y_{i1}}^{Y_{i2}} f(u)\pi^{(2)}(S_i-u) du$, where $Y_{i1}$ and $Y_{i2}$ are the ages of the individual at the beginning and the end of the calendar year recalled by the individual.
\item[{\sc Case}] (v):  When $\delta_i=1$ and $\varepsilon_i=3$ (the event has occurred but the $i$th individual cannot recall the time at all), the contribution of the individual to the likelihood is $\int_{0}^{S_i}f(u)\pi^{(3)}(S_i-u) du$.
\end{description}
Therefore, the overall likelihood is
\begin{align}
\prod_{i=1}^{n}[\bar{F}(S_i)]&^{1-\delta_i}
\Biggl[\left(f(T_i)\pi^{(0)}(S_i-T_i)\right)^{I_{(\varepsilon_i=0)}} \left(\int_{M_{i1}}^{M_{i2}}f(u)\pi^{(1)}(S_i-u) du\right)^{I_{(\varepsilon_i=1)}} \times \notag \\ &\left(\int_{Y_{i1}}^{Y_{i2}}f(u)\pi^{(2)}(S_i-u) du\right)^{I_{(\varepsilon_i=2)}} \left(\int_{0}^{S_i} f(u)\pi^{(3)}(S_i-u) du\right)^{I_{(\varepsilon_i=3)}}\Biggr]^{\delta_i}.
\label{ourM1}
\end{align}
%
%
%
Note that when 
$\pi^{(1)}=\pi^{(2)}=0$, the likelihood \eqref{ourM1} reduces to \eqref{MirSen}. When $\pi^{(1)}=\pi^{(2)}=0$ and $\pi^{(0)}$ is a constant, it becomes a constant multiple of the likelihood corresponding to non-informatively interval censored data. If $\pi^{(0)}=\pi^{(1)}=\pi^{(2)}=0$, it reduces to the current status likelihood \eqref{YN}.

While the proposed model is specific to the data at hand, it can easily be adjusted for arbitrary types of recall, which need not even be ordered.


The factors in the product likelihood (\ref{ourM1}) have different forms in different cases. We now show that they can be expressed as the common density of some random vector with respect to a suitable dominating measure.


The main challenge to obtaining a common format of the data lies in the fact that
$M_{i1}$, $M_{i2}$, $Y_{i1}$ and $Y_{i2}$ are the ages of the $i$th individual at specified calendar times. We make use of the fact that these observables are functions of $T_i$ and the date of birth of the $i$th individual. Specifically, for the $i$th subject, let $m_i$ be the serial number of the month of birth within the year of birth and $d_i$ be the time (measured in years) from the beginning of the month of birth till the event of birth.
For the sake of simplicity, we assume that every year has duration $1$ and every month has duration $1/12$.

When $\epsilon_i=1$, i.e., the month of the event is recalled, we write
\begin{equation}
 \begin{array}{l}
M_{i1} = \lfloor 12(d_i + T_i) \rfloor / 12-d_i, \\
M_{i2} = M_{i1}+1/12,
\end{array} \label{Month}
\end{equation}
where $\lfloor\cdot\rfloor$ is the integer part of its argument. Thus, the variables $\lfloor 12(d_i + T_i) \rfloor$, $m_i$ and $d_i$ can be obtained from $M_{i1}$, $M_{i2}$, $m_i$ and $d_i$ and vice versa. Likewise, when $\epsilon_i=2$, i.e., the year of the event is recalled, we write
\begin{equation}
 \begin{array}{l}
Y_{i1}=\lfloor \big(T_i + d_i + (m_i-1)/12 \big) \rfloor-\big(d_i+(m_i-1)/12\big),\\
Y_{i2} = Y_{i1}+1.
\end{array} \label{Year}
\end{equation}
It is clear that the variables $\lfloor \big(T_i + d_i + (m_i-1)/12 \big) \rfloor$, $m_i$ and $d_i$ are equivalent to $Y_{i1}$, $Y_{i2}$, $m_i$ and $d_i$. Therefore, we define the variable
\begin{equation}
V_i= \left\{
\begin{array}{ll}
T_i & \text {if $\varepsilon_i=0$ , $\delta_i=1$,}\\
\lfloor 12(d_i + T_i) \rfloor / 12 & \text {if $\varepsilon_i=1$ , $\delta_i=1$,}\\
\lfloor \big(T_i + d_i + (m_i-1)/12 \big) \rfloor & \text {if $\varepsilon_i=2$ , $\delta_i=1$,}\\
0 & \text {if $\varepsilon_i=3$, $\delta_i=1$, or if $\delta_i=0$,}
\end{array} \right. \label{W}
\end{equation}
which captures the essential part of the occasionally observable variables $T_i$, $M_{i1}$, $M_{i2}$, $Y_{i1}$ and $Y_{i2}$, and subsequently work with the observable vector
\begin{equation}
 Y_i=(S_i,V_i,\varepsilon_i,\delta_i,m_i,d_i).  \label{Y}
\end{equation}

We have already assumed that the $T_i$s (time-to-event) are samples from the distribution $F$ and the $S_i$s (ages on interview date) are samples from another distribution. We now denote by $G_1$, $G_2$ and $G_3$ the distributions of $S_i$, $m_i$ and $d_i$, respectively, for every $i$. The distribution $G_2$ is defined over the set $\{1,2,\ldots, 12\}$, and $G_3$ is defined over the interval $[0,1/12]$. The latter assumption disregards the fact that $d_i$ is known only up to days (measured as fixed fractions of a year), to keep the description simple.


Theorem~\ref{Thm1} presented below gives the density of~$Y_i$, after the subscript $i$ is dropped for simplicity. The dominating probability measure used for defining this density is $\mu=\vartheta_1\times\vartheta_2\times\vartheta_3\times\vartheta_4\times\vartheta_5\times\vartheta_6$ where $\vartheta_1$ is the measure with respect to which $G_1$ has a density (e.g., the counting or the Lebesgue measure, depending on whether $G_1$ is discrete or continuous), $\vartheta_2$ is the sum of the counting and the Lebesgue measures, each of $\vartheta_3,\vartheta_4$ and $\vartheta_5$ is the counting measure and $\vartheta_6$ is the Lebesgue measure \citep{Ash_2000}.

\begin{theorem}\label{Thm1}
The density of $Y=(S,V,\varepsilon,\delta,m,d)$ with respect to the measure $\mu$ is
\begin{eqnarray}
\mbox{}\hskip-10pt&&\hskip-25pt h(s,v,\varepsilon,\delta,m,d)\nonumber\\
\mbox{}\hskip-10pt&=&\left\{
\begin{array}{ll}
g_1(s)g_2(m)g_3(d)\bar{F}(s) & \text {if $\delta=0$},\\
g_1(s)g_2(m)g_3(d)f(v)\pi^{(0)}(s-v)I_{(v<s)} & \text {if $\varepsilon=0$ and $\delta=1$},\\
g_1(s)g_2(m)g_3(d)\int_{v-d}^{min(s,v+\frac{1}{12}-d} f(u)\pi^{(1))}(s-u) du & \text {if $\varepsilon=1$ and $\delta=1$},\\
g_1(s)g_2(m)g_3(d)\int_{v-d-\frac{m-1}{12}}^{min(s,v+1-d-\frac{m-1}{12})} f(u)\pi^{(2)}(s-u) du & \text {if $\varepsilon=2$ and $\delta=1$},\\
g_1(s)g_2(m)g_3(d)\int_{0}^{s} f(u)\pi^{(3)}(s-u) du & \text {if $\varepsilon=3$ and $\delta=1$},
\end{array} \right. \label{densitY}
\end{eqnarray}
where $g_1$, $g_2$ and $g_3$ are the densities of $G_1$, $G_2$ and $G_3$ with respect to the measures $\vartheta_1$, $\vartheta_5$ and $\vartheta_6$, respectively.
\end{theorem}

Theorem~\ref{Thm1} implies that the likelihood \eqref{ourM1} can be written as
\begin{align}
&\prod_{i=1}^{n}[\bar{F}(S_i)]^{1-\delta_i}
\Biggl[\left(f(V_i)\pi^{(0)}(S_i\!-\!V_i)\right)^{I_{(\varepsilon_i=0)}} \!\left(\int_{V_i-d_i}^{V_i-d_i+\frac{1}{12}}\!\!f(u)\pi^{(1)}(S_i\!-\!u) du\!\right)^{I_{(\varepsilon_i=1)}} \notag \\ &  \times\! \left(\int_{V_i-d_i-\frac{m_i-1}{12}}^{V_i-d_i-\frac{m_i-1}{12}+1}\!\!f(u)\pi^{(2)}(S_i\!-\!u) du\!\right)^{I_{(\varepsilon_i=2)}} \!\!\!\left(\int_{0}^{S_i} \!\!f(u)\pi^{(3)}(S_i\!-\!u) du\!\right)^{I_{(\varepsilon_i=3)}} \Biggr]^{\delta_i}\!\!, \notag \\
& \qquad \qquad =\frac{\prod_{i=1}^{n}h(S_i,V_i,\varepsilon_i,\delta_i,m_i,d_i)}{\prod_{i=1}^{n}g_1(S_i)g_2(m_i)g_3(d_i)},
\label{likedensity}
\end{align}
where the $i$th factor is the conditional density of $(V_i,\varepsilon_i,\delta_i)$ given $(S_i,m_i,d_i)$.

\section{Parametric estimation}\label{Param}
Suppose the forms of the functions $\bar{F}$, $f$, $\pi^{(0)}$, $\pi^{(1)}$, $\pi^{(2)}$ and $\pi^{(3)}$ in the likelihood~\eqref{ourM1} are known up to a few parameters, and accordingly they are written as $\bar{F}_\theta$, $f_\theta$, $\pi^{(0)}_{\eta}$, $\pi^{(1)}_{\eta}$, $\pi^{(2)}_{\eta}$ and $\pi^{(3)}_{\eta}$, respectively. The MLE of the (possibly vector) parameters $\theta$ and $\eta$ are obtained by maximizing~\eqref{ourM1}.



Since the equivalent likelihood (\ref{likedensity}) is identified as a product of conditional densities, standard results for consistency and asymptotic normality of the MLE become applicable. The regularity conditions for these results would then be specified in terms of the density of $Y_i$. In the first section of the supplementary material, we provide easily verifiable sufficient conditions that involve the density $f_\theta$ (the density of $T_i$) and the functions $\pi_{\eta}^{(0)},\pi_{\eta}^{(1)},\pi_{\eta}^{(2)}$ and $\pi_{\eta}^{(3)}$, which define the conditional probability distribution of the random variable $\varepsilon_i$ given $T_i$ and $S_i$.

\section{Non-parametric estimation} \label{nonparametric}
\subsection{Non-parametric MLE} \label{NPMLE}
Before embarking on the task of estimation, we establish the following result on the issue of identifiability.
\begin{theorem}\label{Thm2}
The distribution functions $G_1$ and $F$, and recall probabilities $\pi_{}^{(k)}$, $k=0,1,2,3$ are identifiable from $h$ in \eqref{densitY}.
\end{theorem}
The likelihood \eqref{ourM1} is difficult to maximize because of the
integrals contained in the expression. In order to simplify it, we assume that the function $\pi^{(l)}$ in \eqref{ourM1} is piecewise constant, having the form
$\pi_{}^{(l)}(x)=b_{l1} I(x_1<x\leq x_2)+b_{l2}I(x_2<x\leq x_3)+\ldots+b_{lL}I(x_L<x <\infty)$, $l=0,1,2,3$,
where $0=x_1<x_2<\cdots < x_L$ are a chosen set of time-points and $b_{l1},b_{l2},\ldots,b_{lL}$ are unspecified parameters taking values in the range $[0,1]$ such that $\sum_{l=0}^{3} b_{lj}=1$ for $j=1,2,\ldots,L$. Then the likelihood (\ref{ourM1}) reduces to
\begin{align}
L=&\prod_{i=1}^{n}[\bar{F}(S_i)]^{1-\delta_i}
\Biggl[\left\{f(T_i) \left(\sum_{l=1}^L b_{0l} I\big(W_{l+1}(S_i)<T_i\leq W_l(S_i)\big)\right)\right\}^{I_{(\varepsilon_i=0)}} \notag \\ &\hskip-20pt \times\!\left\{\sum_{\substack{l=1 \\ [W_{l+1}(S_i),W_l(S_i)]\cap [M_{i1},M_{i2}]\ne \phi }}^L b_{1l} \Big(F\big(\min(W_l(S_i),M_{i2})\big)\!-\!F\big(\max(W_{l+1}(S_i),M_{i1})\big)\Big)\right\}^{I_{(\varepsilon_i=1)}} \notag \\
&\hskip-20pt \times\!\left\{\sum_{\substack{l=1 \\ [W_{l+1}(S_i),W_l(S_i)]\cap [Y_{i1},Y_{i2}]\ne \phi }}^L b_{2l} \Big(F\big(\min(W_{l}(S_i),Y_{i2})\big)\!-\!F\big(\max(W_{l+1}(S_i),Y_{i1})\big)\Big)\right\}^{I_{(\varepsilon_i=2)}} \notag \\
&\hskip-20pt \times\!\left\{\sum_{l=1}^L b_{3l} \Big(F(W_l(S_i))-F(W_{l+1}(S_i))\Big)\right\}^{I_{(\varepsilon_i=3)}}\Biggr]^{\delta_i},  \label{ourMINT}
\end{align}
where $W_l(S_i)=(S_i-x_l)\vee t_{min}$ for $l=1,\ldots,L$ and
$W_{L+1}(S_i)=t_{min}, i=1,2,\ldots,n$.
The likelihood \eqref{ourMINT}
involves probabilities assigned to
intervals of the type $[t,t_{max}]$ or $(t,t_{max}]$, as per the
baseline probability distribution. Since these intervals have
overlap, we try to write them as unions of some disjoint intervals.
Let ${\cal I}_1$, ${\cal I}_2$, ${\cal I}_3$, ${\cal I}_4$ and ${\cal I}_5$ be sets of indices
$i$ (between 1 and $n$) that satisfy the conditions $\delta_i=0$,
$\delta_i\varepsilon_i=1$, $\delta_i(1-\varepsilon_i)=1$, $\delta_i\varepsilon_i=2$ and $\delta_i\varepsilon_i=3$.
respectively. Consider the intervals
\begin{equation}\begin{array}{r@{\hskip3pt}c@{\hskip3pt}ll}
A_i&=&(S_i,t_{max}] & \mbox{for }i\in{\cal I}_1,\\[.4ex]
A_i&=&[T_i,t_{max}] & \mbox{for }i\in {\cal I}_2,\\[.4ex]
A_i'&=&(T_i,t_{max}]& \mbox{for }i\in{\cal I}_2,\\[.5ex]
A_{il}&=& \left\{ \begin{matrix}(W_l(S_i),t_{max}],& l=1,\ldots,L,\\
[W_l(S_i),t_{max}],&l=k+1,\ \ \ \,\\ \end{matrix} \right.&   \mbox{for}\ i\in{\cal I}_2 \cup{\cal I}_3,\\[2ex]
B_{il}&=&[W_{l+1}(S_i) \vee M_{i1},W_{l+1}(S_i) \wedge M_{i1}] & \mbox{for }i\in {\cal I}_4\ \&  \ l=1,\ldots,L,\\[.4ex]
C_{il}&=&[W_{l+1}(S_i) \vee Y_{i1},W_{l+1}(S_i) \wedge Y_{i1}] & \mbox{for }i\in {\cal I}_5\ \& \ l=1,\ldots,L,\\
\end{array}\label{Ai}
\end{equation}
and the sets
\begin{equation}\begin{array}{r@{\hskip3pt}c@{\hskip3pt}l}
{\cal A}_1&=&\{A_i: \ \ i\in{\cal I}_1\},\\[.25ex]
{\cal A}_2&=&\{A_i\setminus A_i': \ \ i\in {\cal I}_2\},\\[.25ex]
{\cal A}_3&=&\{A_i':\ \ i\in{\cal I}_2\},\\[.25ex]
{\cal A}_4&=&\{A_{i(l+1)}\setminus A_{il}:\ \ 1\le l\le L \mbox{ and}\ i\in{\cal I}_3\},\\[.25ex]
{\cal A}_5&=&\{ B_{il}:\ \ 1\le l\le L \mbox{ and}\  i\in{\cal I}_4\},\\[.25ex]
{\cal A}_6&=&\{ C_{il}:\ \ 1\le l\le L \mbox{ and}\  i\in{\cal I}_5\}.\\
\end{array}\label{CalA}
\end{equation}
As $F$ is absolutely continuous, the elements
of ${\cal A}_2$ and ${\cal A}_3$ are distinct with probability~1.
Let $n_i$ be the cardinality of ${\cal I}_i$,
$i=1,2,3,4,5$.
We arrange the singleton elements of ${\cal A}_2$ in increasing order, and
denote them as $B_1, B_2,\ldots, B_{n_2}$. We also arrange the elements
of ${\cal A}_3$ in the corresponding order and denote them as
$B_{n_2+1},B_{n_2+2},\ldots,B_{2n_2}$. We then collect the unique
elements of ${\cal A}_1\cup {\cal A}_4\cup {\cal A}_5\cup {\cal A}_6$ that are distinct from
$B_1,B_2,\ldots,B_{2n_2}$, and denote them as
$B_{2n_2+1},B_{2n_2+2},\ldots, B_M$. Observe that the collection
$B_1,B_2,\ldots,B_M$ consists of the distinct elements of $\bigcup_{i=1}^{6}{\cal A}_i$, arranged in a
particular order. Denote the non-empty subsets of the index set
$\{1,2,\ldots,M\}$ by $s_1,s_2,\ldots,s_{2^M-1}$. Define
\begin{equation}
I_r=\left\{\bigcap_{i\in
s_r}B_i\right\}\bigcap\left\{\bigcap_{i\notin
s_r}B_i^c\right\}\qquad \mbox{for }r=1,2,\ldots,2^M-1.\label{Ir}
\end{equation}
Some of the $I_r$s may be empty sets, denoted here by $\phi$. Let
\begin{eqnarray}
{\cal C}&=&\{s_r:\,I_r\ne\phi,\,1\le r\le 2^M-1\},  \label{C}\\
{\cal A}&=&\{I_r:\,I_r\ne\phi,\,1\le r\le 2^M-1\}.  \label{A}
\end{eqnarray}
It can be verified that the elements of ${\cal A}$ are distinct and
disjoint.

Note that each of the intervals $B_1,\ldots,B_M$ is a union of
disjoint sets that are members of ${\cal A}$. For any Borel set $A$,
suppose $P(A)$ is the probability assigned to $A$ as per the
probability distribution~$F$. Let $p_r=P(I_r)$, for $I_r \in \cal A$. Then the
likelihood \eqref{ourMINT} reduces to
 \begin{align}
 L=& \prod_{i\in {\cal I}_1} \!\!\left(\sum\limits_{\substack{r:I_r \subseteq A_i \\ s_r\in {\cal C} }} \!\!p_r\!\right)\!\times\!\!
 \prod_{i\in {\cal I}_2}\!\!\left(\!1-\sum_{l=1}^{L} (b_{1l}+b_{2l}+b_{3l}) I_{(T_i \in A_{i(l+1)}\backslash A_{il})}\!\right)\!\cdot\! \left(\sum\limits_{\substack{r:I_r \subseteq A_{i}\backslash A_{i'}\\ s_r\in {\cal C}}}\!\!p_r\!\right)\notag\\
 &\hskip-20pt \times\! \prod_{i\in {\cal I}_3}\!\left[\sum_{l=1}^{L} b_{3l}\!\!\left(\!\sum\limits_{\substack{r:I_r \subseteq A_{i(l+1)}\backslash A_{il}\\ s_r\in {\cal C}}}\!\!p_r\!\right)\!\right]\!\times\! \prod_{i\in {\cal I}_4}\!\left[\sum\limits_{\substack {l=1 \\ [W_{l+1}(S_i),W_l(S_i)]\cap [M_{i1},M_{i2}]\ne \phi }}^{L} \!\!b_{1l}\!\!\left(\sum\limits_{\substack{r:I_r \subseteq B_{il}\\ s_r\in {\cal C}}}\!\!p_r\!\right)\!\right]\notag\\
 &\hskip-20pt \times \prod_{i\in {\cal I}_5}\left[\sum\limits_{\substack {l=1 \\ [W_{l+1}(S_i),W_l(S_i)]\cap [Y_{i1},Y_{i2}]\ne \phi }}^{L} b_{2l}\!\left(\sum\limits_{\substack{r:I_r \subseteq C_{il}\\ s_r\in {\cal C}}}\!p_r\!\right)\right]. \label{ourM_int1}
\end{align}
Thus, maximizing the likelihood \eqref{ourMINT} amounts
to maximizing \eqref{ourM_int1} with respect to $p_r$
for $s_r \in \cal C$.

There is a partial order among the members of ${\cal C}$ in the
sense that some sets are contained in others.
Consider the following subsets of ${\cal C}$.
\begin{eqnarray}
{\cal C}_1&=&\{s:\,s\in{\cal C};\ \mbox{there is another element $s'\in{\cal C}$, such that $s\subset s'$}\}, \nonumber\\
{\cal C}_2&=&\{s:\,s\in{\cal C};\ \mbox{there is another element $s'\in{\cal C}$, such that} \nonumber\\
           && s'\backslash(s\cap s') \mbox{ consists of a singleton $j$ and }s\backslash(s\cap s')=\{j+n_2\}\}, \nonumber\\
{\cal C}_0&=&{\cal C}\backslash({\cal C}_1\cup{\cal C}_2).
\label{C0}
\end{eqnarray}
We now present a result which shows that maximization of the likelihood can be
restricted to ${\cal C}_0$.
\begin{theorem}\label{Thm3}
Maximizing the likelihood \eqref{ourM_int1} with respect to $p_r$
for $s_r \in \cal C$ is equivalent to maximizing it with respect to
$p_r$ for $s_r \in {\cal C}_0$, i.e.,
\begin{equation*}
    \mathop{\max\limits_{p_r:p_r
\in [0,1], \sum_{s_r \in \cal C}p_r=1}} L\ \ =\ \
\displaystyle\mathop{\max\limits_{p_r:p_r \in [0,1], \sum_{s_r \in
{\cal C}_0}p_r=1}} L
\end{equation*}
\end{theorem}

It transpires from the above theorem that
the likelihood has the same maximum value, irrespective of whether $s_r$ is chosen from the class ${\cal C}$ or ${\cal C}_0$. Therefore,
we can replace ${\cal C}$ by ${\cal C}_0$ in \eqref{ourM_int1}.

Let us relabel the intervals $I_j,$ $s_j\in{\cal C}_0$, by
$J_1,J_2,\ldots,J_\nu$. Further, let ${\cal A}_0=\{J_1,J_2,\ldots,J_\nu\}$ and $q_j=P(J_j)$ for
$j=1,2,\ldots,\nu$. If the likelihood \eqref{ourM_int1} is rewritten with the condition $s_r\in{\cal C}$ replaced by the equivalent condition $I_r\in{\cal A}$, then Theorem~\ref{Thm3} shows that the latter condition can be replaced by $I_r\in{\cal A}_0$. In other words, maximizing the likelihood
\eqref{ourM_int1} is equivalent to maximizing
\begin{align}
&L(p,\eta)\notag\\
&=
\prod_{i\in {\cal I}_1}\!\! \left(\sum_{j:J_j \subseteq A_i } \!\!q_j\!\!\right)\!\!\times\!\!
\prod_{i\in {\cal I}_2}\!\!\left(\!1-\sum_{l=1}^{L} (b_{1l}+b_{2l}+b_{3l}) I_{(T_i \in A_{i(l+1)}\backslash A_{il})}\!\right)\!\cdot\! \left(\sum_{j: J_j \subseteq A_{i}\backslash A_{i'}}\!\!q_j\!\!\right)\notag\\
 &\times\!\! \prod_{i\in {\cal I}_3}\!\!\left[\sum_{l=1}^{L} \!b_{3l}\!\!\left(\sum_{j: J_j\subseteq A_{i(l+1)}\backslash A_{il}}\!\!q_j\!\!\right)\!\!\right]\!\!\times\!\! \prod_{i\in {\cal I}_4}\!\!\left[\sum\limits_{\substack {l=1 \\ [W_{l+1}(S_i),W_l(S_i)]\cap [M_{i1},M_{i2}]\ne \phi }}^{L} \! b_{1l}\!\!\left(\sum_{j:J_j \subseteq B_{il}}\!\!q_j\!\right)\!\!\right]\notag\\
 &\times \prod_{i\in {\cal I}_5}\left[\sum\limits_{\substack {l=1 \\ [W_{l+1}(S_i),W_l(S_i)]\cap [Y_{i1},Y_{i2}]\ne \phi }}^{L} b_{2l}\left(\sum_{j:J_j \subseteq C_{il}} q_j\right)\right]=\prod_{i=1}^n\left(\sum_{j=1}^v \alpha_{ij}q_j\right),\label{ourMRed}
\end{align}
with respect to the vector parameters $p=(q_1,q_2,\ldots,q_\nu)^T$ and $\eta=(b_{11},\ldots,$ $b_{1L},b_{21},\ldots,b_{2L},b_{31},\ldots,b_{3L})^T$,
subject to the restrictions $\sum_{j=1}^\nu q_j=1$, $0\le q_1,\ldots,q_\nu\leq 1$, where
\begin{equation}
 \alpha_{ij} = \left\{
 \begin{array}{ll}
 I_{(J_j\subseteq A_{i})} & \mbox{if}~~ i\in{\cal I}_1,\\
 \left(1-\sum_{l=1}^{L} (b_{1l}+b_{2l}+b_{3l}) I_{(T_i \in A_{i(l+1)}\backslash A_{il})}\right).I_{(J_j\subseteq A_{i}\backslash A'_{i})} & \mbox{if}~~ i\in {\cal I}_2,\\
 \sum_{l=1}^{L} b_{1l} .I_{(J_j\subseteq A_{i(l+1)}\backslash A_{il})} & \mbox{if}~~i\in {\cal I}_3,\\
 \sum_{l=1}^{L} b_{2l} .I_{(J_j\subseteq B_{il})} & \mbox{if}~~i\in {\cal I}_4,\\
 \sum_{l=1}^{L} b_{3l} .I_{(J_j\subseteq C_{il})} & \mbox{if}~~i\in {\cal I}_5,\\
\end{array} \right.    \label{alpha}
\end{equation}
for $i=1,\ldots,n,$ and $j=1,\ldots,\nu$.

Now consider the set ${\cal A}_2=\left\{\{T_i\},\ i\in{\cal I}_2\right\}$ defined in
\eqref{CalA}, with cardinality set~$n_2$ (defined after \eqref{CalA}). The task of maximization is
simplified further through the following result, which is
interesting by its own right.
\begin{theorem}\label{Thm4}
The set ${\cal A}_2$ is contained in the set ${\cal A}_0$ almost
surely. Further, if $G$ is a discrete distribution with finite support, then the probability of ${\cal A}_0$ being equal to ${\cal
A}_2$ goes to one as $n\rightarrow \infty$.
\end{theorem}
We are now ready for the next result regarding the existence and
uniqueness of the NPMLE. The uniqueness is established probabilistically under the condition that
$n_2$, the number of cases with exact recall, goes to infinity.

\begin{theorem}\label{Thm5}
The likelihood \eqref{ourMRed} has a maximum. Further, if $G$ is a discrete distribution with finite support, then the
probability that it has a unique maximum goes to one, as
$n_2\rightarrow \infty$.
\end{theorem}

\subsection{Self-consistency approach for estimation} \label{selfconsistency}
We follow the footsteps of \cite{Efron_1967}
and \cite{Turnbull_1976}
to obtain the NPMLE through the self consistency approach.
For $\ i=1,2,\ldots, n,$ let
$$
 L_{ij} = \left\{
 \begin{array}{ll}
 1 & \mbox{if}~~ T_i\in J_j,\\
 0 &\mbox{otherwise},\\
 \end{array} \right.
$$
When $i\in {\cal I}_2$, the value of $L_{ij}$ is known.
If $i\notin {\cal I}_2$, its expectation with respect to the probability vector
$p$ is given by
\begin{equation}
E(L_{ij})=\frac{\alpha_{ij}q_j}{\sum\limits_{j=1}^\nu
\alpha_{ij}q_j}=\mu_{ij}(p),\hspace{.4cm}    \mbox{say.}
\label{mu1}
\end{equation}
Thus, $\mu_{ij}(p)$ represents the probability that the $i$-th observation lies in $J_j$.
The average of these probabilities across the $n$ individuals,
\begin{equation}
 \frac{1}{n}\sum\limits_{i=1}^n\mu_{ij}(p)= \mathcal{\pi}_j(p),\hspace{.4cm}    \mbox{say,}  \label{pi}
\end{equation}
should indicate the probability of the interval $J_j$. Thus, it is
reasonable to expect that the vector $p$ would satisfy the
equation
\begin{equation}
q_j=  \mathcal{\pi}_j(p) \hspace{.4cm} \mbox{for}\quad 1\leq j \leq \nu.  \label{sc}
 \end{equation}
An estimator of $p$ may be called self consistent if it satisfies \eqref{sc}. The form of these equations suggests the following iterative procedure.
\begin{description}
\item[{\sc Step I.}] Obtain a set of initial estimates $q^{(0)}_j\hspace{.2cm} (1\leq j\leq m)$.
\item[{\sc Step II.}] At the $n$th stage of iteration, use current estimate, ${p}^{(n)}$, to evaluate $\mu_{ij}(p^{(n)})$ for $ i=1,2,\ldots, n,\ j=1,2,\ldots, \nu$ and $\mathcal{\pi}_j({p}^{(n)})$ for $j=1,2,\ldots, \nu$ from \eqref{mu1} and \eqref{pi}, respectively.
\item[{\sc Step III.}] Obtain updated estimates ${p}^{(n+1)}$ by setting $q^{(n+1)}_j=  \mathcal{\pi}_j({p}^{(n)})$.
\item[{\sc Step IV.}] Return to Step II with ${p}^{(n+1)}$ replacing ${p}^{(n)}$.
\item[{\sc Step V.}] Iterate; stop when the required accuracy has been achieved.
\end{description}

The following theorem shows that equation \eqref{sc} defining a
self consistent estimator must be satisfied by an NPML estimator of
$p$.
\begin{theorem}\label{Thm6}
An NPML estimator of ${p}$ must be self consistent.
\end{theorem}

\subsection{A computationally simpler estimator} \label{Simple}
The computational complexity of the NPMLE depends on the number of
segments $(k)$ used in the piecewise constant formulation of the
function $\pi_\eta$.
One can conceive of a computational simplification
on the basis of Theorem~\ref{Thm3}. According to this theorem, the NPMLE has
mass only at points of exact recall of the event, when $n$ is large.
In such a case, the likelihood \eqref{ourMRed} involves $J_j$s that
are singletons only.

Formally, let $t_1,\ldots,t_{n_2}$ be the ordered set of distinct ages
at event that have been perfectly recalled, and $q_1^*,\ldots,q_{n_2}^*$
be the probability masses allocated to them. The likelihood
\eqref{ourMRed}, subject to the constraint that $q_j=0$ whenever
$J_j \notin {\cal A}_2$, is equivalent to the unconstrained
maximization~of
\begin{equation}
L(p^*,\eta)=\prod_{i=1}^{n} \left[\sum_{j=1}^{n_2} \alpha_{ij}q^*_j \right], \label{aproxour}
\end{equation}
with respect to the parameters $p^*=(q^*_1,\ldots,q^*_{n_2})^T$ and  $\eta$,
over the set
$$\Re^*=\left\{(p^*,\eta) |\sum_{j=1}^{n_2} q^*_j=1,\quad
0\le q^*_1,\ldots,q^*_{n_2}\leq 1,\ 0\le b_1\le\cdots\le b_k\le 1
\right\}.$$
Let the likelihood~\eqref{aproxour} be maximized at
$(\hat p^*,\hat{\eta}^*)$, where $\hat p^*=(\hat{q}^*_1,\ldots,\hat{q}^*_{n_2})^T$. We define an
approximate NPMLE (AMLE) of $F$ as
\begin{equation}
 \tilde{F}_n(t) = \sum_{j: t_j\leq t} \hat{q}^*_j.  \label{Ftilde}
\end{equation}
Both NPMLE and AMLE depend on $L$, the number of line segments in the descriptions of recall probabilities. One can use successively higher values of $L$ (e.g., higher powers of 2) and choose a value after which further increase does not add substantially to the details. A data analytic illustration of this principle in given in Section~\ref{DataAnalysis}.

\subsection{Consistency of the estimators}\label{Consistency}
Let $\Theta$ be the set of all distribution functions over $[t_{min},t_{max}]$, i.e.,
\begin{align}
\Theta =&\{F\,:\,[t_{min},t_{max}]\rightarrow[0,1];\,F\mbox{ right continuous, nondecreasing};\\ \nonumber
&\hskip2.1inF(t_{min})=0;\,F(t_{max})=1\}.
\end{align}
and $\overline{\Theta}$ be the set of all sub-distribution
functions, i.e.,
\begin{align}
\overline{\Theta} =&\{F\,:\,[t_{min},t_{max}]\rightarrow[0,1];\,F\mbox{ right continuous, nondecreasing};\\ \nonumber
&\hskip2.1inF(t_{min})=0;\,F(t_{max})\le1\}.
\end{align}
Note that, with respect to the topology of vague convergence,
$\overline{\Theta}$ is compact by Helley's selection theorem.
Further, let $F_0$ denote the true distribution of the time of
occurrence of landmark events with density $f_0$, and $F_0(t_{min})=0$.

For any given distribution $F\in\Theta$ having masses restricted to
the set $\{t_1,\ldots,t_{n_2}\}$, the log of the likelihood
(\ref{aproxour}) can be rewritten as a function of $F$ (instead $q^*_1,\ldots,q^*_{n_2}$) as
\begin{equation}
\ell(F)=\sum_{i=1}^{n} \log\left[\sum_{j=1}^{n_2}
\alpha_{ij}\left\{F(t_j)-F(t_{j^-})\right\} \right].
\label{ouraprox2}
\end{equation}
Define the set
\begin{equation}
 {\cal E}=\{F\,:\,F\in \Theta,\,E[\ell(F)-\ell(F_0)]=0\},  \label{class}
\end{equation}
which is an equivalence class of the true distribution $F_0$.

Strong consistency of the AMLE and weak consistency of the NPMLE are established by the following
theorems.
\begin{theorem}\label{Thm7}
In the above set-up, the AMLE $\{\tilde{F_n}\}$ converges
almost surely to the equivalence class ${\cal E}$ of the true
distribution $F_0$, in the topology of vague convergence.
\end{theorem}
\begin{theorem}\label{Thm8}
{In the set-up described before Theorem~\ref{Thm7}, the NPMLE $\{\hat{F_n}\}$ converges in
probability to the equivalence class ${\cal E}$ of the true
distribution $F_0$, in terms of the L\'{e}vy distance.}
\end{theorem}

\section{Simulation of performance}
\label{Simulation}
\subsection{Parametric estimation} \label{parsim}
We consider the MLEs based on the current status likelihood (\ref{YN}) (described here as Current Status MLE), the likelihood \eqref{MirSen} based on binary recall (described here as Binary Recall MLE) and the likelihood (\ref{ourM1}) based on partial recall (described here as Partial Recall MLE). Computation of the three MLEs is done through numerical optimization of likelihood using the Quasi-Newton method \cite{Nocedal_2006}.

For the purpose of simulation, we generate samples of time-to-event from the Weibull distribution with shape and scale parameters $\theta_1$ and $\theta_2$, respectively. Thus, $\theta=(\theta_1,\theta_2)$. We generate the recall probabilities through the multinomial logistic model,
$\log\Big(\pi_{\eta}^{(k)}(u)/\pi_{\eta}^{(0)}(u)\Big)=\alpha_k+\beta_ku$, $k=1,2,3$. Since $\sum_{k=0}^{3}\pi_{\eta}^{(k)}(u)=1,$ the probabilities can be written as
\begin{equation}
 \begin{array}{l}
\pi_{\eta}^{(0)}(u)=1/\big(1+\sum_{k=1}^{3}e^{\alpha_k+\beta_ku}\big),\\
\pi_{\eta}^{(k)}(u)=e^{\alpha_k+\beta_ku}/\left(1+\sum_{k=1}^{3}e^{\alpha_k+\beta_ku}\right), \quad \ k=1,2,3,
\end{array} \label{pifuncfinal}
\end{equation}
where $\eta=(\alpha_1,\alpha_2,\alpha_3,\beta_1,\beta_2,\beta_3)$. Further, we generate the `age at interview' from the discrete uniform distribution over [8,21].

We use the following sets of values of the parameters.
\begin{enumerate}
\item[(i)] $\theta=(10,12)$ and $\eta=(-0.05,-0.05,-0.05,0.01,0.01,0.01)$,
\item[(ii)] $\theta=(10,12)$ and $\eta=(-2,-1,-0.4,0.05,0.3,0.02)$,
\item[(iii)] $\theta=(10,12)$ and $\eta=(-2,-0.7,-1,0.5,0.06,0.2)$,
\item[(iv)] $\theta=(10,12)$ and $\eta=(-2,-2,-2,0.3,0.08,0.08)$.
\end{enumerate}
Note that for the chosen value of $\theta$, the median of the Weibull distribution turns out to be 11.6, which is in line with the median estimated from the data described in Section~\ref{intro} under a simpler model \citep{Mirzaei_2015}. Also, the chosen values of $\eta$ correspond to the following probabilities of different types of recall, five years after the event.
\begin{enumerate}
\item[(i)] $\pi_{\eta}^{(0)}(5)=\pi_{\eta}^{(1)}(5)=\pi_{\eta}^{(2)}(5)=\pi_{\eta}^{(3)}(5)=0.25$,
\item[(ii)] $\pi_{\eta}^{(0)}(5)=0.28$, $\pi_{\eta}^{(1)}(5)=0.46, \pi_{\eta}^{(2)}(5)=0.21, \pi_{\eta}^{(3)}(5)=0.05$,
\item[(iii)] $\pi_{\eta}^{(0)}(5)=0.23$, $\pi_{\eta}^{(1)}(5)=0.15, \pi_{\eta}^{(2)}(5)=0.23, \pi_{\eta}^{(3)}(5)=0.38$,
\item[(iv)] $\pi_{\eta}^{(0)}(5)=0.5$, $\pi_{\eta}^{(1)}(5)=0.1, \pi_{\eta}^{(2)}(5)=0.1, \pi_{\eta}^{(3)}(5)=0.3$.
\end{enumerate}
Choice (iv) is meant to favour the Binary Recall MLE, as the chances of partial recall are slim. Choice (ii) should favour the Partial Recall MLE. Choice (iii), with a high probability attached to `no recall', gives Current Status MLE its best chance. Choice (i) does not favour any single method.

While computing the Binary Recall MLE, we assume the following form of the non-recall probability function $\pi_\eta$:
$$\log\Big(\pi_\eta(u)/1-\pi_\eta(u)\Big)=\alpha+\beta u.$$

We run 1000 simulations for each of the above combinations of parameters, for sample size $n=100,300,1000$, to compute the empirical bias, the standard deviation (Stdev) and the mean squared error (MSE) for the MLEs of the parameter $\theta=(\theta_1,\theta_2)$, the median time-to-event, and $\pi_{\eta}^{(0)}(5)$ (the exact recall probability 5~years after the event), based on the three likelihoods. These indicators of performance, for the combinations of parameter values given in case (i) to case (iv), are reported in Table~\ref{t:tableone} for $n=100$.

\begin{table}
\centering
\caption{Bias, standard deviation (Stdev) and MSE of estimated parameters for $n=100$}
\label{t:tableone}
\scalebox{0.9}{
\begin{tabular}{ccc@{\hskip3pt}c@{\hskip3pt}cc@{\hskip3pt}c@{\hskip3pt}cc@{\hskip3pt}c@{\hskip3pt}c}
\\[-1.5ex]
\hline
Case	&	Param	&\multicolumn{3}{c}{Current Status MLE}&\multicolumn{3}{c}{Binary Recall MLE}&\multicolumn{3}{c}{Partial Recall MLE}\\
\cline{3-11}
	&		&	Bias	&	Stdev	&	MSE	&	Bias	&	Stdev	&	 MSE	&	Bias	&	Stdev	&	MSE	\\
\hline
(i)	    &	$\theta_1$	&	1.698	&	5.368	&	31.67	&	0.487	 &	1.701	&	3.127	&	0.247	&	1.07	&	1.207	\\
	    &	$\theta_2$	&	-0.071	&	0.329	&	0.113	&	-0.023	 &	0.233	&	0.055	&	-0.01	&	0.165	&	0.027	\\
	    &	Median	    &	-0.047	&	0.338	&	0.116	&	-0.012	 &	 0.241	&	0.058	&	-0.003	&	0.172	&	0.029	\\
	    &	$\pi_{\eta}^{(0)}(5)$	&	-	&	-	&	-	&	-0.004	 &	0.054	&	0.003	&	0.0001	&	0.054	&	0.002	\\
\hline
(ii)	&	$\theta_1$	&	1.845	&	5.270	&	31.15	&	0.520	 &	1.745	&	3.314	&	0.214	&	0.952	&	0.952	\\
	    &	$\theta_2$	&	-0.058	&	0.341	&	0.119	&	-0.01	 &	0.341	&	0.051	&	-0.011	&	0.145	&	0.021	\\
	    &	Median   	&	-0.031	&	0.347	&	0.121	&	-0.002	 &	 0.240	&	0.058	&	-0.005	&	0.152	&	0.023	\\
	    &	$\pi_{\eta}^{(0)}(5)$	&	-	&	-	&	-	&	-0.018	 &	0.057	&	0.004	&	0.0007	&	0.053	&	0.003	\\
\hline
(iii)	&	$\theta_1$	&	1.930	&	5.091	&	29.63	&	0.573	 &	1.828	&	3.669	&	0.381	&	1.322	&	1.893	\\
	    &	$\theta_2$	&	-0.07	&	0.331	&	0.114	&	-0.024	 &	0.243	&	0.059	&	-0.007	&	0.182	&	0.033	\\
	    &	Median	    &	-0.037	&	0.337	&	0.115	&	-0.011	 &	 0.255	&	0.065	&	-0.002	&	0.193	&	0.037	\\
	    &	$\pi_{\eta}^{(0)}(5)$	&	-	&	-	&	-	&	-0.026	 &	0.056	&	0.004	&	-0.003	&	0.060	&	0.004	\\
\hline
(iv)	&	$\theta_1$	&	1.803	&	5.333	&	31.66	&	0.262	 &	1.291	&	1.735	&	0.253	&	1.146	&	1.377	\\
	    &	$\theta_2$	&	-0.062	&	0.332	&	0.114	&	-0.018	 &	0.191	&	0.04	&	-0.014	&	0.174	&	0.031	\\
	    &	Median   	&	-0.036	&	0.340	&	0.117	&	-0.012	 &	 0.202	&	0.041	&	-0.008	&	0.185	&	0.034	\\
	    &	$\pi_{\eta}^{(0)}(5)$	&	-	&	-	&	-	&	-0.012	 &	0.064	&	0.004	&	0.001	&	0.067	&	0.004	\\
\hline
\end{tabular}
}
\end{table}

In cases (i)--(iii), it is found that the bias and the standard deviation (and consequently the MSE) of the Partial Recall MLE is generally less than (and sometimes comparable to) those of the other two estimators and its performance improves with increasing sample size. The Current Status MLE, which uses the least amount of information from the data, has the poorest performance even in case (iii), where a substantial proportion of the subjects are designed to have no recollection of the event date. The substantial gap between the performance of the Binary Recall MLE and the Partial Recall MLE shows that the later estimator is able to utilize the additional information available from partial recall data. Similar tables for $n=300$ and 1000 are given in the supplementary material, to save space. The conclusions are similar, though all the methods perform better when the sample size increases.

For sensitivity analysis, we consider the following mixture model for the time-to-event distribution \begin{equation*}
    \gamma \log \mbox{Normal} + (1- \gamma) \mbox{Weibull},
\end{equation*}
with the parameters of $\log$ Normal  $(\mu=2.45, \sigma^2=0.07)$ and $\gamma=0.2$ and 0.5. Note that for the chosen values of $\mu$ and $\sigma^2$, the median of the time-to-event distribution remains 11.6. The rest of the simulation set-up also remain the same as before.
The sensitivity analysis is done for the sample size of $n=300$ with 1000 simulations runs, under the assumption $\gamma=0$, and reported in the supplementary material. The summary of the findings is that the miss-specification does not alter the relative order of the performances of the three estimators when $\gamma=0.2$. When $\gamma=0.5$, Partial Recall MLE has smaller MSE than Binary Recall MLE, as before, but both of these estimators are outperformed by the Current Status MLE.

\subsection{Non-parametric estimation} \label{nonparsim}
We generate sample times-to-event ($T$) from the Weibull distribution with shape and scale parameters $\theta=(10,12)$ as before, but truncate the generated samples to the interval [8,16]. This truncated distribution has median of 11.6. The corresponding `time of interview' ($S$) is generated from the discrete uniform distribution over $\{8,\ldots,21\}$. These choices are in line with the data set described in Section~\ref{intro}, and lead to about 29\%\ cases of no-occurrence of the event till the time of interview ($S<T$). As for the recall probabilities, we use \eqref{NPMLE} with $L=4$, $x_1=0$, $x_2=3$, $x_3=6$, $x_4=9$ and three sets of values of the parameters, described bellow.
\begin{description}
 \item Case (a) $b_0=(0.15,0.10,0.08,0.05)$, $b_1=(0.28,0.2,0.15,0.1)$, $b_2=(0.22$, $0.25,0.17,0.1)$, $b_3=(0.35,0.45,0.6,0.75)$, which correspond to overall probabilities of exact recall $E[\pi^{(0)}(S{-}T)|S{>}T]=0.10$, recall up to calendar month $E[\pi^{(1)}(S{-}T)|S{>}T]=0.20$, recall up to calendar year  $E[\pi^{(2)}(S{-}T)|S{>}T]=0.20$ and no recall $E[\pi^{(3)}(S{-}T)|S{>}T]=0.50$.
 \item Case (b) $b_0=(0.69,0.55,0.49,0.31)$, $b_1=b_2=(0.08,0.05,0.03,0.02)$, $b_3=(0.15,0.35,0.45,0.65)$ which correspond to overall exact recall probability 0.55, calendar month recall probability 0.05, calendar year recall probability 0.05 and no-recall probability 0.35.
 \item Case (c) $b_0=1-(b_1+b_2+b_3)$, $b_1=b_2=b_3=(0.25,0.25,0.25,0.25)$, which correspond to equal probability (0.25) of each type of recall.
\end{description}


It has been observed by \cite{Mirzaei_2016} that in the special case of binary recall, the performances of AMLE and NPMLE are comparable. Therefore, we choose not to run simulation for NPMLE, which involves heavier computation. Instead, we compare the performance of the AMLE estimated from \eqref{Ftilde} (described here as Partial Recall AMLE) with those of the AMLE based on \eqref{MirSen}, proposed by \cite{Mirzaei_2016} (described here as Binary Recall AMLE), and the empirical estimate of $F$ (described here as EDF). The EDF is used only as a hypothetical benchmark of performance that could have been achieved with complete data.

The Partial Recall AMLE is implemented by using the correct value of
$L,x_1,x_2,\ldots$, $x_L$ in \eqref{NPMLE}, while
the likelihood~\eqref{aproxour} is recursively maximized alternately with respect to the probability parameter $p^*$ and the nuisance parameter $\eta=(b_0,b_1,b_2,b_3)^T$.

\begin{figure}
\hspace*{- 25mm}
  \includegraphics[height=8in,width=7.5in]{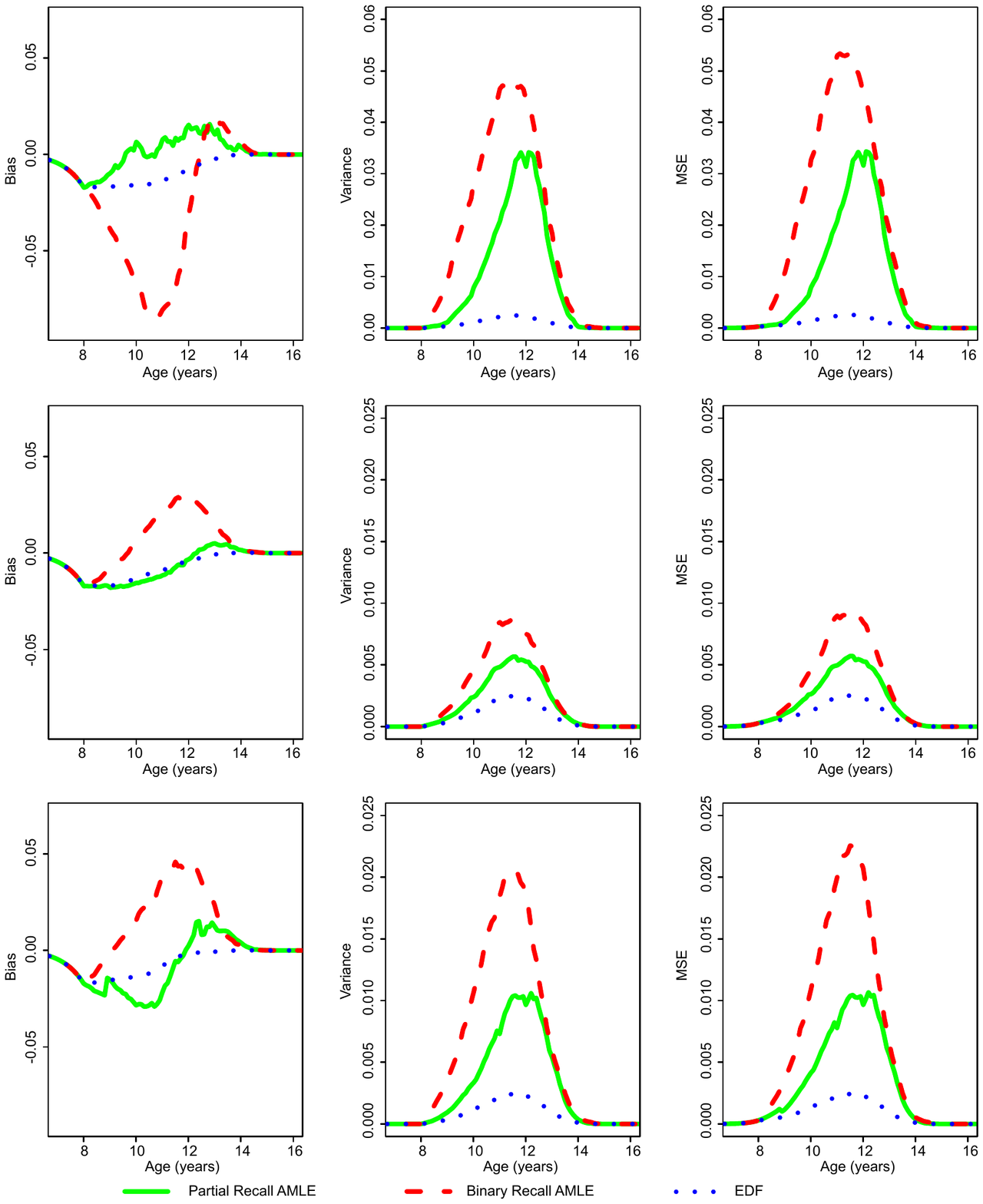}
\caption[]{Comparison of bias, variance and MSE of the estimator for $n=100$ in cases (a) (top panel), (b) (middle panel) and (c) (bottom panel)} \label{fig_2}
\end{figure}

Figure~\ref{fig_2} shows plots of the bias, the variance and the mean square error (MSE) of the three estimators for different ages, when $n=100$ and parameters of the recall functions \eqref{NPMLE} are chosen as in Cases (a), (b) and~(c). The Partial Recall AMLE is found to have smaller bias, variance and MSE than the Binary Recall AMLE, although its performance is expectedly poorer than that of the EDF.

Plots similar to Figure~\ref{fig_2} for $n=300$ and 1000 are given in the supplementary material. At those sample sizes, the performance parameters of partial AMLE are found to be closer to those of EDF than those of binary AMLE.

\section{Adequacy of the Model} \label{adequacy}
One can use the chi-square goodness of fit test to check how well the assumed parametric model actually fits the data. For this purpose, the data may be transformed to the vector $Y=(S,V,\varepsilon,\delta,m,d)$, and the support of the distribution of this vector may be appropriately partitioned, depending on the availability of data. An example is given in the next section.

Modeling of the recall probability functions is a critical issue. One has to choose suitable functional forms, and also strike a balance between a flexible model and a parsimonious one with fewer parameters. We provide below an exploratory technique for selecting the functional forms.

As we have seen in Section~\ref{nonparametric}, use of the piecewise constant form \eqref{NPMLE} of the recall probabilities reduces the likelihood (\ref{ourM1}) to the likelihood \eqref{ourMINT}.
If the distribution of $T$ is known, one can obtain the MLE of the parameters $ b_{l1}, b_{l2} ,\ldots , b_{lk},\ l=0,1,2,3$.
The conditional MLE of the piecewise constant functions $\pi_{}^{(1)},\pi_{}^{(2)},\pi_{}^{(3)}$ and $\pi_{}^{(0)}$, for any given $F_\theta$ can be obtained iteratively. By using a candidate parametric form $\pi_{\eta}^{(1)},\pi_{\eta}^{(2)},\pi_{\eta}^{(3)}$ and $\pi_{\eta}^{(0)}$, one can first estimate the MLEs $\hat{\theta}$ and $\hat\eta$ and then compare the plots of $\hat\pi_{\eta}^{(1)},\hat\pi_{\eta}^{(2)},\hat\pi_{\eta}^{(3)}$ and $\hat\pi_{\eta}^{(0)}$ with the plots of the conditional MLE of the piecewise constant versions of $\pi_{}^{(1)},\pi_{}^{(2)},\pi_{}^{(3)}$ and $\pi_{}^{(0)}$, with $F_\theta$ held fixed at $F_{\hat{\theta}}$. An example of this graphical check is given in the next section.

In addition, comparative plots of $F_{\hat{\theta}}$ computed for an assumed form of the recall probability functions and the piecewise constant forms mentioned in Section~\ref{NPMLE}, can also serve as a graphical check of the adequacy of that assumed form. An example of this graphical check for the data set of next section is given in the supplementary material.

\section{Data Analysis}
\label{DataAnalysis}
For the data set described in Section~\ref{intro}, the landmark event is the onset of menarche in young and adolescent females.
In a parametric analysis, we used the Weibull model for menarcheal age and the multinomial logistic model for the recall probabilities $\pi_{\eta}^{(0)}, \pi_{\eta}^{(1)}, \pi_{\eta}^{(2)}$ and $\pi_{\eta}^{(3)}$, as in Section~\ref{parsim}.
 We used the three different methods mentioned in Section~\ref{parsim} for estimating the parameters $\theta_1$ and $\theta_2$ as well as the median of the age at menarche.
Table~\ref{t:tabletwo} gives a summary of the findings. The Partial Recall MLEs have smaller standard errors than those of the other two estimators.
\begin{table}[t!]
 \centering
 \def\~{\hphantom{0}}

  \caption{Different estimates of parameters for the menarcheal data}
\label{t:tabletwo}
\scalebox{1.0}{
\begin{tabular}{p{4.5cm}c c c}
\\[-1.5ex]
     \noalign{\hrule height 1pt}
  Estimator &$\theta_1$ (Stdev) & $\theta_2$ (Stdev) & Median (Stdev)\\
         \noalign{\hrule height 1pt}
  Current Status MLE    & 19.05 (5.31) & 11.65 (0.20) &11.42 (0.043)\\
     Binary Recall MLE  & 10.32 (0.91) & 12.27 (0.15) &11.84 (0.025)\\
  Partial Recall MLE    & 9.432 (0.61) & 12.25 (0.12) &11.78 (0.010)\\
 \noalign{\hrule height 1pt}
\end{tabular}
}
\end{table}

\begin{figure}
\centering
\includegraphics[height=4in,width=3in,angle=-90]{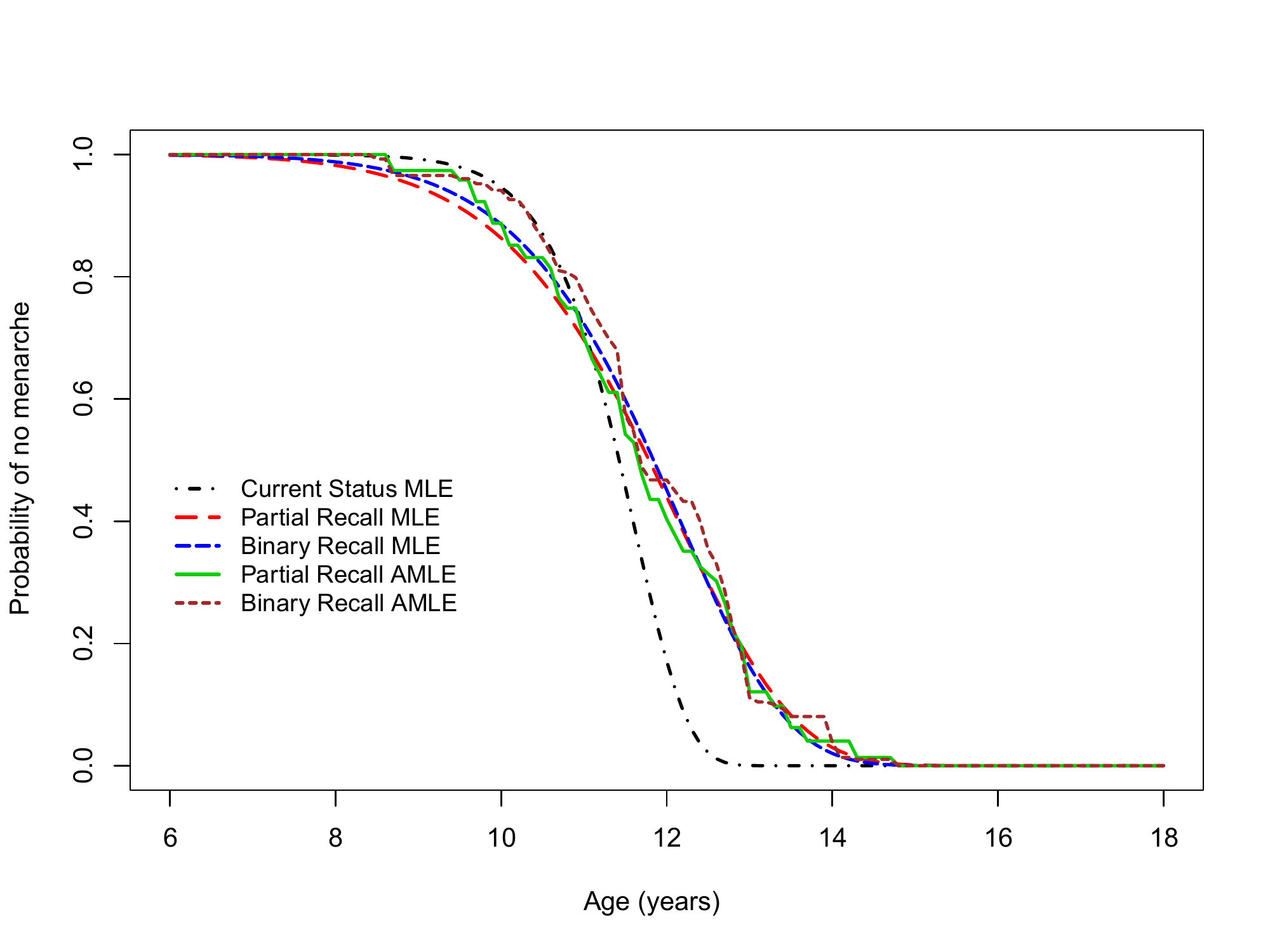}
\caption[]{Survival functions for the menarcheal data based on five methods} \label{fig_3}
\end{figure}
Figure~\ref{fig_3} shows the survival functions estimated from the three parametric methods, the Partial Recall AMLE presented in Section~\ref{Simple} (with knot points of the recall probability functions chosen as in the first paragraph of Section~\ref{nonparsim}) and Binary Recall AMLE (with the same knot points). The parametric MLEs are not very far from the non-parametric AMLEs. Though there appears to be little difference between the Partial Recall and Binary Recall MLEs, their standard errors are different (check Figure~3 of supplementary material).

In order to formally check how well the assumed parametric model fits the data, we use the chi-square goodness of fit test, by discretizing the range of the hexatuple $(S,V,\varepsilon,\delta,m,d)$.
The range of $S$ is split into the intervals $[7,14]$ and $(14,21]$, the range of $d$ is split into the intervals $[0,1/24]$ and $(1/24,1/12]$, while the range of $V$ is split into the sets $[0,11.84]$ and $(11.84,21]$ ($11.84$ being the median of the observed non-zero values of $V$). The ranges of $\varepsilon$ and $\delta$ have four points ($0$, $1$, $2$ and $3$) and two points ($0$ and $1$), respectively, none of which are clubbed. The range of $m$ is the interval $[0,11]$, which is not split.
When $\delta=0$, the value of $\varepsilon$ is irrelevant and $V=0$, i.e., there are four bins for the two groups of values of $S$ and two groups of $d$. When $\delta=1$ and $\varepsilon=3$, $V$ can only be zero and again there are only four bins. When $\delta=1$ and $\varepsilon=0,1$ or 2, in each case there are eight bins arising from two groups of values of $S$ and two groups of non-zero values of $V$ and $d$. Thus, we have a total of 32 bins.

In order to avoid small expected frequency in some cells we merge some bins where expected frequency is less than $5$. After this pruning, we have a reduced total of~21~bins.~There~are~8 parameters to estimate. Thus, the null distribution should be $\chi^2$ with 12 degrees of freedom. The p--value of the test statistic for the given data happens to be 0.169. Therefore, violation of the chosen model is not indicated.

\begin{figure}
\centering
 \includegraphics[height=4in,width=3in,angle=-90]{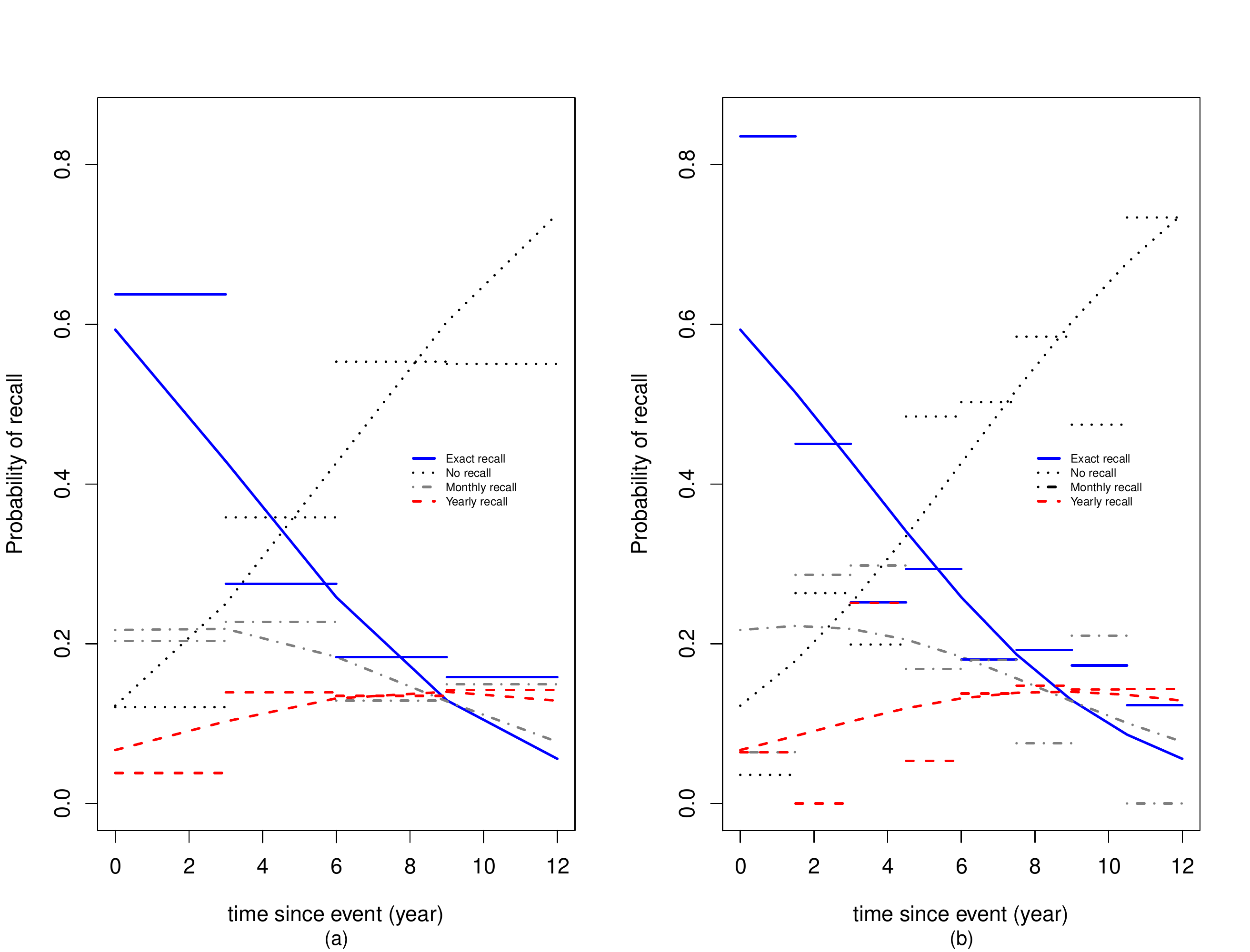}
\caption[]{Comparison of estimated logistic recall probabilities with estimated piecewise constant recall probabilities with (a) 4 pieces, (b) 8 pieces} \label{fig_4}
\end{figure}

We now check the adequacy of the functional form of the $\pi_{\eta}^{(l)}$s by comparing the $\pi_{\hat\eta}^{(k)}$s with the conditional MLE of the corresponding piecewise constant function in~\eqref{NPMLE}, as indicated in the last section.
For the given data, the largest value of $S_i-T_i$ in a perfectly recalled case happens to be 10.88 years. Therefore, we consider recall functions over the interval 0 to 12 years. With $F$ chosen as Weibull and $\theta_1$ and $\theta_2$ fixed at the values reported in the last row of Table~\ref{t:tabletwo}, we obtained the conditional MLE of the values of $\pi_{\eta}^{(0)}$, $\pi_{\eta}^{(1)}$, $\pi_{\eta}^{(2)},\pi_{\eta}^{(3)}$ in different segments of equal length. Figure~\ref{fig_4}(a) shows the plots of the estimated recall probabilities under the logistic and the piecewise constant models, with number of segments $L=4$. The estimated functions are found to be close to each other for $l=0,1,2,3$. Figure~\ref{fig_4}(b) shows the same plots for $L=8$. The finer partition seems unnecessary.
As another check of the functional form of the recall probability, we estimated the survival functions of time-to-event from the proposed parametric method using the multiple logistic regression model presented in Section~\ref{parsim}  and the piecewise constant recall probability model introduced in Section~\ref{NPMLE} (with knot points of the recall probability functions chosen as in the first paragraph of Section~\ref{nonparsim}). Figure~4 of supplementary material shows the two estimates of the survival function, which happen to be very close to each other.

\begin{figure}[tbh]
\centering
 \includegraphics[height=3.6in,width=2.8in,angle=-90]{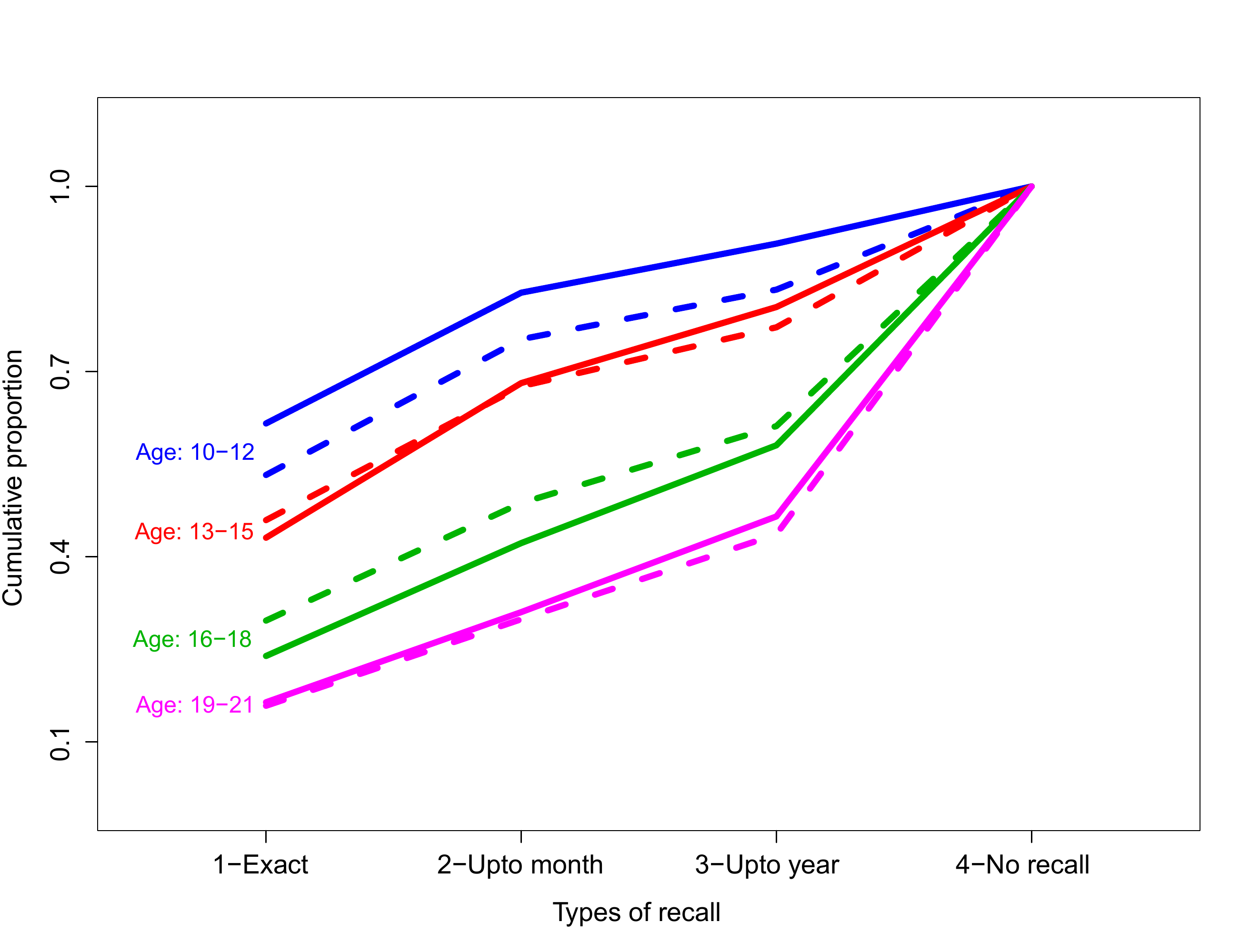}
\caption[]{\small Cumulative proportions (solid lines) and model based estimated probabilities (dashed lines) of decreasing degrees of recall in menarcheal data} \label{fig_5}
\end{figure}

We have seen the cumulative proportions of decreasing degrees of recall for different age ranges in the case of the menarcheal data in Figure~\ref{fig_1}. As an additional check for the assumed model, we consider the model based estimates of these cumulative proportions for ages $s=11,14,17$ and 20 (i.e., at the middle of the respective age intervals). We used the Partial Recall MLE of parameters $\hat\theta$ and $\hat\eta$ to calculate $f_{\hat\theta}$ and $\pi_{\hat\eta}^{(j)}$ for $j=0,1,2,3$ and then computed the requisite probabilities through numerical integration. Figure~\ref{fig_5} shows the cumulative proportions in different age groups
(solid lines) along with the corresponding model based estimates (dashed lines). The estimated
probabilities are quite close to the empirical proportions.

\vspace{-1cm}
\section{Concluding Remarks}
\label{s:discuss}
The aim of this paper has been to offer a realistic model for time-to-event based on partial recall information through an informative censoring model, where the range of relevant dates may depend on calendar time (rather than time elapsed since the event). The simulations and the data analysis of the menarcheal data set show that there is much to be gained from
partial recall information in the form of the event falling in a calendar month or a calendar year. Many other forms of partial recall information may be handled in a similar way. As the simulations reported in Section~\ref{Simulation} show, a particular category of partial recall (eg. recall up to a calendar month or year) is justified if that category is not very rare in the data.

The recalled time-to-event can sometimes be erroneous. Grouping of the uncertainly recalled event date by the calendar month or year may reduce the error to some extent. If one adopts this solution, the method presented in this paper provides a viable method of analysis. \cite{Skinner_1999}, while working with data without instances of non-recall, has modeled erroneously recalled time-to-event as $t'_i=t_ik_i$, where $t_i$ is the correct time-to-event and $k_i$ is a multiplicative error of recall that is independent of $t_i$. Since $k_i$s are unobservable, they have used a mixed-effects regression model to account for erroneous recalls. One may investigate whether a similar adjustment in the term $f_{\theta}(T_i)$ of the likelihood~\eqref{ourM1}, improves the analysis.

The Cox regression model has been adapted to the retrospective recall model for binary recall data \citep{MirSen_2014}, and an adaptation to partial recall would be interesting. The multiple logistic regression model provides a framework for incorporating covariate effect on the recall probabilities also. These problems will be taken up in future.
%
\section{Software}
\label{sec9}
Software in the form of R code, together with the data set and complete documentation is available at GitHub (\url{https://github.com/rahulfrodo/PartialRecall}).
\section{Supplementary Material}
\label{sec10}
Supplementary material is available online at
\url{http://biostatistics.oxfordjournals.org}.
\section {Acknowledgements}
This research is partially sponsored by the project ``Physical growth, body composition and nutritional status of the Bengal school aged children, adolscents, and young adults of Calcutta, India: Effects of socioeconomic factors on secular trends'', funded by the Neys Van Hoogstraten Foundation of the Netherlands. The authors thank Professor Parasmani Dasgupta of the Biological Anthropology Unit of ISI, for making the data available for this research. The authors thank an anonymous referee and an associate editor for suggesting useful changes that improved the content and the presentation of the paper.

\bibliographystyle{biorefs}
\bibliography{refs}

\end{document}


\begin{center}
  {\bf Supplementary Materials}  
\end{center}
\vskip-.8in
\title{Estimating menarcheal age distribution from partially recalled data}
\author{Sedigheh Mirzaei Salehabadi$^\ast$, Debasis Sengupta, Rahul Ghosal\\[4pt]
}

\markboth%
{S. Mirzaei S. and others}
{Estimating menarcheal age distribution from partially recalled data}

\maketitle
\section{Parametric estimation}
{\bf Regularity conditions}

The following are sufficient conditions for consistency of the MLE of $\theta$ and $\eta$ in the likelihood (2.5) of the main paper, where $F$, $f$, $\pi^{(0)}$, $\pi^{(1)}$, $\pi^{(2)}$ and $\pi^{(3)}$ are replaced by $F_\theta$, $f_\theta$, $\pi^{(0)}_\eta$, $\pi^{(1)}_\eta$, $\pi^{(2)}_\eta$ and $\pi^{(3)}_\eta$, respectively. These conditions are obtained by adapting the conditions of Theorem 7.1.1 of \cite{Lehman_1999}, which apply to the density of $Y_i$, and expressing them in terms of $f_\theta$ (the density of $T_i$) and the functions $\pi^{(0)}_\eta$, $\pi^{(1)}_\eta$, $\pi^{(2)}_\eta$ and $\pi^{(3)}_\eta$, which define the conditional probability distribution of the random variable $\varepsilon_i$ given $T_i$ and $S_i$.

 \begin{enumerate}
\item[(C1)] The parameters $\theta$ and $\eta$ are identifiable with respect to the family of densities $f_{\theta}$ of the time-to-event and the family of functions $\pi_{\eta}^{(k)}$,  $k=1,2,3$. In other words, $f_{\theta_1} = f_{\theta_2}$ implies $\theta_1 =\theta_2$ and congruence of $\pi_{\eta_1}^{(k)}$ and $\pi_{\eta_2}^{(k)}$ for $k=1,2,3$ implies $\eta_1 = \eta_2$.
\item[(C2)] The parameter spaces for $\theta$ and $\eta$ are open.
\item[(C3)] The set $A_1=\left\{t: f_{\theta}(t)>0 \right\}$ is independent of $\theta$ and the set
$A_2=\Bigl\{t: \pi^{(k)}(t) \in (0,1)\Bigr.$, $\Bigl.k=0,1,2,3\Bigr\}$ is independent of $\eta$.
\item[(C4)] The functions $f_\theta(t)$, $\pi^{(1)}(t)$, $\pi^{(2)}(t)$ and $\pi^{(3)}(t)$ are differentiable with respect to $\theta$ and $\eta$ for all $t$ such that the derivative is absolutely bounded by a $\mu$-integrable function.
\end{enumerate}

%

The additional conditions for asymptotic normality of the MLE are conditions 1-5 of Theorem 18 \cite[Chapter 18]{Ferguson_1996}, where the log-likelihood is
\begin{align}
&\hskip-20pt\ell(\theta,\eta)=\nonumber\sum_{i=1}^{n}\biggl[
\delta_i I_{(\varepsilon_i=3)}\log\left(\int_{0}^{S_i} f_{\theta}(u)\pi^{(3)}(S_i-u) du\right)\nonumber\\
&+\delta_i I_{(\varepsilon_i=2)}\log\left(\int_{Y_{i1}}^{Y_{i2}} f_{\theta}(u)\pi^{(2)}(S_i-u) du\right)\biggr.\nonumber\\
&+\biggl.
\delta_i I_{(\varepsilon_i=1)}\log\left(\int_{M_{i1}}^{M_{i2}} f_{\theta}(u)\pi^{(1)}(S_i-u) du\right)\nonumber\\
&+ \delta_i I_{(\varepsilon_i=0)}\log\left(f_{\theta}(T_i)\pi^{(0)}(S_i-T_i))\right)+(1-\delta_i)\log\left(\bar{F}_\theta(S_i)\right)\biggr].  \label{logL}
\end{align}

{\bf Proof of Theorem 2.1}

In the second case, the density can be derived as,
\begin{align*}
&\hskip-15pt h(s,v,0,1,m,d)\\
&= g_1(s)g_2(m)g_3(d)\frac{\partial P(V < v, \delta=1, \varepsilon=1|s,m,d)}{\partial v}\\
 &=g_1(s)g_2(m)g_3(d)\lim_{h\rightarrow0} \frac{P(v<V\leqslant v+h,\delta=1, \varepsilon=1|s,m,d)}{h}\\
 &=g_1(s)g_2(m)g_3(d)\lim_{h\rightarrow0} \frac{P(v<T\leqslant v+h,T\leq s,\varepsilon=1)}{h}\\
 &=g_1(s)g_2(m)g_3(d)\lim_{h\rightarrow0} \frac{P(v<T\leqslant v+h,\varepsilon=1)}{h}\\
 &=g_1(s)g_2(m)g_3(d)\lim_{h\rightarrow0} \frac{E_T[P(v<T\leqslant v+h|T)\pi^{(0)}(s-T)I_{(v\leq s)}]}{h}\\
&=g_1(s)g_2(m)g_3(d)\lim_{h\rightarrow0} \frac{\int_{v}^{v+h}f_{\theta}(u)\pi^{(0)}(s-u)duI_{(v\leq s)}}{h}\\
&=g_1(s)g_2(m)g_3(d)f_{\theta}(v)\pi^{(0)}(s-v)I_{(v\leq s)}.
\end{align*}
The density in the other cases can be obtained by considering the corresponding probability masses:
\small
\begin{align*}
h(s,v,\varepsilon,0,m,d)&=P(V=0,\delta=0|s,m,d)g_1(s)g_2(m)g_3(d)\\&=P(T>S|S=s)g_1(s)g_2(m)g_3(d)=\bar{F}_\theta(s)g_1(s)g_2(m)g_3(d); \\
h(s,v,3,1,m,d)&=E_T[g_1(s)g_2(m)g_3(d)P(T\leq s|T,m,d,s)\pi^{(3)}(s-T)]\\
&=\int_{0}^{s}g_1(s)g_2(m)g_3(d)f_{\theta}(u)\pi^{(3)}(s-u)du\\
&=g_1(s)g_2(m)g_3(d)\int_{0}^{s}f_{\theta}(u)\pi^{(3)}(s-u)du\\
h(s,v,1,1,m,d)&=g_1(s)g_2(m)g_3(d)P(V=v,\varepsilon=1,\delta=1|s,m,d)\\
&=g_1(s)g_2(m)g_3(d)P(\lfloor 12(d + T) \rfloor / 12= v,\varepsilon=1,\delta=1|s,m,d)\\
&=g_1(s)g_2(m)g_3(d)P(12v\leq 12(d + T) < 12v+1,\\
&\hskip30pt \varepsilon=1,\delta=1|s,m,d)\\
&=g_1(s)g_2(m)g_3(d)\int_{v-d}^{v+\frac{1}{12}-d}f_\theta(u)\pi^{(1)}(s-u)du;\\
h(s,v,2,1,m,d)&=g_1(s)g_2(m)g_3(d)P(V=v,\varepsilon=2,\delta=1|s,m,d)\\
&=g_1(s)g_2(m)g_3(d)P(\lfloor \big(T + d + (m-1)/12 \big) \rfloor=v,\\
&\hskip30pt \varepsilon=2,\delta=1|s,m,d)\\
&=g_1(s)g_2(m)g_3(d)P(v-d-(m-1)/12 \leq T\\
&\hskip30pt < v+1-d-(m-1)/12, \varepsilon=2,\delta=1|s,m,d)\\
&=g_1(s)g_2(m)g_3(d)\int_{v-d-\frac{m-1}{12}}^{v+1-d-\frac{m-1}{12}}f_\theta(u)\pi^{(2
)}(s-u)du;
\end{align*}
\normalsize

\section{Non-parametric estimation}
{\bf Proof of Theorem 4.1}

In last theorem it is shown that, the density of $Y=(S,V,\varepsilon,\delta,m,d)$ with respect to the measure $\mu$ is
\begin{eqnarray}
&&\hskip-20pt h(s,v,\varepsilon,\delta,m,d)\nonumber\\
&\hskip-10pt =& \left\{
\begin{array}{ll}
g_1(s)g_2(m)g_3(d)\bar{F}(s) & \text {if $\delta=0$},\\
g_1(s)g_2(m)g_3(d)f(v)\pi^{(0)}(s-v)I_{(v<s)} & \text {if $\varepsilon=0$ and $\delta=1$},\\
g_1(s)g_2(m)g_3(d)\int_{v-d}^{min(s,v+\frac{1}{12}-d)} f(u)\pi^{(1)}(s-u) du & \text {if $\varepsilon=1$ and $\delta=1$},\\
g_1(s)g_2(m)g_3(d)\int_{v-d-\frac{m-1}{12}}^{min(s,v+1-d-\frac{m-1}{12})} f(u)\pi^{(2)}(s-u) du & \text {if $\varepsilon=2$ and $\delta=1$},\\
g_1(s)g_2(m)g_3(d)\int_{0}^{s} f(u)\pi^{(3)}(s-u) du & \text {if $\varepsilon=3$ and $\delta=1$},
\end{array} \right.\nonumber\\
&&\mbox{} \label{densitY}
\end{eqnarray}
where $g_1$, $g_2$ and $g_3$ are the densities of $G_1$, $G_2$ and $G_3$ with respect to the measures $\vartheta_1$, $\vartheta_5$ and $\vartheta_6$, respectively.
\begin{description}
\item (a) We have, from \eqref{densitY}, $ g_1(s)= \int_{v,\varepsilon,\delta,m,d}
h(s,v,\varepsilon,\delta,m,d) $ and hence $G_1$ are identifiable from $h$. It is the same for $g_2$, $G_2$ and $g_3$, $G_3$.

Also we have \[\bar{F}(s)=\frac{h(s,0,\varepsilon,0,m,d)}{g_1(s)g_2(m)g_3(d)},\]  \label{Fbar}
and \[\pi^{(0)}(s-v)=\frac{h(s,v,1,1,m,d)}{g_1(s)g_2(m)g_3(d)},\]  \label{pi1}
that show $F$ and $\pi^{(0)}$ are identifiable form $h$.
 \item (b) For the sake of contradiction, let us assume there are two $\pi^{(1)}$s, say $\pi^1_2$ and $\pi^2_2$, such that their substitution in the right hand side of \eqref{densitY} produces the same function $h_1=h_2=h$. By differentiating $h$ w.r.t. $v$, we get
\begin{align*}
  &\frac{d h(s,v,2,1,m,d)}{d v}\\ &=f\big((v+1/12)-d\big)\pi^1_2\big(s-(v+1/12)+d\big)-f(v-d)\pi^1_2(s-v+d)\\
   &=f\big((v+1/12)-d\big)\pi^2_2\big(s-(v+1/12)+d\big)-f(v-d)\pi^2_2(s-v+d).
\end{align*}
Hence,
\begin{align}
  &f\big((v+1/12)-d\big)\pi^1_2\big(s-(v+1/12)+d\big)-\pi^2_2\big(s-(v+1/12)+d\big)\notag\\
   &=f\big(v-d\big)\big[\pi^1_2\big(s-v+d\big)-\pi^2_2(s-v+d)\big],\notag
\end{align}
which implies,
\begin{align*}
  &\frac{f\big((v+1/12)-d\big)}{f(v-d)}\\
  &=\frac{\pi^1_2\big(s-v+d\big)-\pi^2_2(s-v+d)}{\pi^1_2\big(s-(v+1/12)+d\big)-\pi^2_2\big(s-(v+1/12)+d\big)}>0\quad \forall s.
\end{align*}
Since the numerator and the denominator are the same function evaluated at two different points, we have either \[ \pi^1_2\big(s-v+d\big)-\pi^2_2(s-v+d) >0\quad \forall s,\]
or, \[ \pi^1_2\big(s-v+d\big)-\pi^2_2(s-v+d) <0\quad \forall s.\]
Without loss of generality, let  $\pi^1_2\big(s-v+d\big)-\pi^2_2(s-v+d) >0 \quad \forall s,$ i.e.,\ $\pi^1_2>\pi^2_2$. Since
\begin{align*}
 h_1&= \int_{v-d}^{(v+1/12)-d} f(u) \pi^1_2(s-u) du\notag\\&= \int_{v-d}^{(v+1/12)-d}f(u)\big[ \pi^1_2(s-u)-\pi^2_2(s-u)\big]du\\
 &+ \int_{v-d}^{(v+1/12)-d}f(u)\pi^1_2(s-u)du,
\end{align*}
we have
\[ h_1-h_2= \int_{v-d}^{(v+1/12)-d}f(u)\big[ \pi^1_2(s-u)-\pi^2_2(s-u)\big]du >0, \]
which contradicts the assumption. Therefore, $\pi^{(1)}$ is uniquely defined for any given $h$. A similar argument can be used to show that $\pi^{(2)}$ is identifiable from~$h$.
From the identity $\sum_{k=0}^{3}\pi_k=1$, we conclude that all the $\pi$s are identifiable from $h$.
\end{description}

{\bf Proof of Theorem 4.2}

From the definitions of $\cal C$ and ${\cal C}_0$, we
can rewrite the likelihood (4.20) as follows.
\small
 \begin{align}
 L=& \prod_{i\in {\cal I}_1} \left(\sum\limits_{\substack{r:I_r \subseteq A_i \\ s_r\in {\cal C}\backslash{\cal C}_0 }} p_r+\sum\limits_{\substack{r:I_r \subseteq A_i \\ s_r\in {\cal C}_0 }} p_r\right)\notag\\
 &\times
 \prod_{i\in {\cal I}_2}\left(1-\sum_{l=1}^{L} (b_{1l}+b_{2l}+b_{3l}) I_{(T_i \in A_{i(l+1)}\backslash A_{il})}\right)\cot \left(\sum\limits_{\substack{r:I_r \subseteq A_{i}\backslash A_{i'}\\ s_r\in {\cal C}\backslash{\cal C}_0 }}p_r+\sum\limits_{\substack{r:I_r \subseteq A_{i}\backslash A_{i'}\\ s_r\in {\cal C}_0 }}p_r\right)\notag\\
 &\times \prod_{i\in {\cal I}_3}\left[\sum_{l=1}^{L} b_{3l}\left(\sum\limits_{\substack{r:I_r \subseteq A_{i(l+1)}\backslash A_{il}\\ s_r\in {\cal C}\backslash{\cal C}_0}}p_r+\sum\limits_{\substack{r:I_r \subseteq A_{i(l+1)}\backslash A_{il}\\ s_r\in {\cal C}_0}}p_r\right)\right]\notag\\ &\times \prod_{i\in {\cal I}_4}\left[\sum\limits_{\substack {l=1 \\ [W_{l+1}(S_i),W_l(S_i)]\cap [M_{i1},M_{i2}]\ne \phi }}^{L} b_{1l}\left(\sum\limits_{\substack{r:I_r \subseteq B_{il}\\ s_r\in {\cal C}\backslash{\cal C}_0}}p_r+\sum\limits_{\substack{r:I_r \subseteq B_{il}\\ s_r\in {\cal C}_0}}p_r\right)\right]\notag\\
 &\times \prod_{i\in {\cal I}_5}\left[\sum\limits_{\substack {l=1 \\ [W_{l+1}(S_i),W_l(S_i)]\cap [Y_{i1},Y_{i2}]\ne \phi }}^{L} b_{2l}\left(\sum\limits_{\substack{r:I_r \subseteq C_{il}\\ s_r\in {\cal C}\backslash{\cal C}_0}}p_r+\sum\limits_{\substack{r:I_r \subseteq C_{il}\\ s_r\in {\cal C}_0}}p_r\right)\right]. \label{ourM_int1}
\end{align}
\normalsize
For any $ s_r\in{\cal C}\backslash {\cal C}_0 $, let ${\cal
A}_r=\{I_{r'}: s_{r'}\in {\cal C}_0, s_r\subset s_{r'} \}$. By definition of
${\cal C}_0$, ${\cal A}_r$ is a non-empty set. The elements of
${\cal A}_r$ are disjoint sets consisting of unions of intervals,
which are subsets of $[t_{min},t_{max}]$. Let $I_{r^*}$ be that
member of ${\cal A}_r$ which satisfies the condition `there is
$\alpha \in I_{r^*}$ such that $\alpha < \beta$ whenever $\beta \in
I_{r^\dag}$ for any $I_{r^\dag} \in {\cal A}_r$' (in some sense, it is the minimal element in ${\cal A}_r$). We are going to show that by
moving mass from any $I_{r}$ to $I_{r^*}$, there won't be
reduction in the contribution of any individual to the likelihood (20). The change in the likelihood would be through the sets $B_j$ such that $j\in (s_{r^*}\backslash s_r)$.

We shall check the effect of shift of mass on the contribution of each individual $(i=1,\ldots,n)$ to the likelihood.

\vskip.1in\noindent Case (a). For any $j\in (s_{r^*}\backslash s_r)$, let $i_j$ be such that $B_j=A_{i_j}$ for $i_j\in {\cal I}_1 $. Since $I_{r^*} \subseteq A_{i_j}$, but $I_{r} \not\subset A_{i_j}$, contribution of individual $i_j$ will increase by shifting mass from $I_r$ to $I_{r^*}$.
\vskip.1in\noindent Case (b). For any $j\in (s_{r^*}\backslash s_r)$, let $i_j$ be such that $B_j=A_{i_j}\backslash A'_{i_j}$ for $i_j\in {\cal I}_2 $. Since $I_{r^*} \subseteq A_{i_j}$, but $I_{r} \not\subset A_{i_j}$, by construction $B_{n_2+j}=A'_{i_j}$ which is disjoint from $B_j$, and we have $n_2+j \notin s_{r^*}$, i.e.,  $n_2+j \notin s_{r}$. This implies $I_r \not\subset B_j$ and $I_r \not\subset B_{n_2+j}$, i.e., $I_{r} \not\subset A_{i_j}$ and $I_{r} \not\subset A'_{i_j}$. Therefore, contribution of individual $i_j$ will increase by shifting mass from $I_r$ to $I_{r^*}$.
\vskip.1in\noindent Case (c). For any $j\in (s_{r^*}\backslash s_r)$, let $i_j$ be such that $B_j=A_{i_j(l+1)}\backslash A'_{i_jl}$ for $i_j\in {\cal I}_3, l=1,2,...,k $. Contribution of individual $i_j$ will increase by shifting mass from $I_r$ to $I_{r^*}$ because $I_{r^*} \in A_{i_j(l+1)}$ and $I_{r^*} \in A_{i_jl}$, but $I_{r}$ is not in either of them.
\vskip.1in\noindent Case (d). For any $j\in (s_{r^*}\backslash s_r)$, let $i_j$ be such that $B_j=B_{i_jl}$ for $i_j\in {\cal I}_4, l=1,2,...,k $. Contribution of individual $i_j$ will increase by shifting mass from $I_r$ to $I_{r^*}$ because $I_{r^*} \in B_{i_jl}$, but $I_{r} \notin B_{i_jl}$.
\vskip.1in\noindent Case (e). For any $j\in (s_{r^*}\backslash s_r)$, let $i_j$ be such that $B_j=C_{i_jl}$ for $i_j\in {\cal I}_5, l=1,2,...,k $. Contribution of individual $i_j$ will increase by shifting mass from $I_r$ to $I_{r^*}$ because $I_{r^*} \in C_{i_jl}$, but $I_{r} \notin C_{i_jl}$.
\vskip.1in It follows that maximizing $L$ can be restricted to
$\{p_r:s_r \in {\cal C}_0\}$.

\bigskip
{\bf Proof of Theorem 4.3}

Let $i \in {\cal I}_2$ and index $j_i$ be such that $S_{j_i}=\{j: T_i \in B_j\}$. Since time-to-event has absolutely continuous distribution, the recalled times $T_i$ for $i \in {\cal I}_2$ are distinct with probability 1. Therefore $T_i \in \{B_1,B_2, \ldots, B_{n_2}\}$ almost surely and $I_{j_i}=T_i$ with probability 1. Also, $S_{j_i}=\{j: T_i \in B_j\}\in {\cal C}_0$. Therefore, ${\cal A}_2\subseteq{\cal A}_0$.

The interview times are discrete valued with finite
domain; $x_1,x_2,\ldots,x_k$ are also finite. So, there are finite number of sets in the form of $A_{il}, B_{il}, C_{il}$. Therefore, even when
$n$ is large, there is at most a finite number (say $N$) of distinct
sets of the form
$$A_s=\left\{\bigcap_{i\in s}B_i\right\}\bigcap\left\{\bigcap_{i\in {{\cal I}_1 \cup {\cal I}_3 \cup {\cal I}_4 \cup {\cal I}_5 \backslash s}}B_i^c\right\},$$
where $s\subseteq {\cal I}_1 \cup {\cal I}_3 \cup {\cal I}_4 \cup {\cal I}_5.$

Let $s^{(1)},
s^{(1)},\ldots,s^{(N)}$ be the index sets corresponding to the $N$
distinct sets described above.
Consider a member of ${\cal A}_0$, say $I_s$, where $s$ is a subset
of $\{1,2,\ldots,n\}$. If $s\subseteq{\cal I}_2$, then it is already
a singleton. If not, it can be written as $s^{(j)}\cup(s\backslash
s^{(j)})$, with $s^{(j)}\subseteq {\cal I}_1 \cup {\cal I}_3 \cup {\cal I}_4 \cup {\cal I}_5$ and
$s\backslash s^{(j)}\subseteq{\cal I}_2$ for some $j \in
\{1,2,\ldots,N\}$. Let us consider further cases.

\vskip0.1in\noindent Case (a). Let $s=s^{(j)} \cup \{r\}$ for $r \in
{\cal I}_2$. In this case, $I_s$ is either a singleton or a null
set. If it is a null set, then it cannot be a member of ${\cal A}$,
and hence of ${\cal A}_0$. Thus, Case (a) contributes only
singletons to ${\cal A}_0$.

\vskip0.05in\noindent Case (b). Let $s=s^{(j)} \cup
\{r_1,r_2,\ldots,r_p\},$ for $r_1,r_2,\ldots, r_p \in {\cal I}_2$
when $p >1$. In this case, $I_s$ is either a singleton or a null set.
Since the absolute continuity of the time-to-event distribution
almost surely precludes coincidence of two sample values (say,
$T_{r_1}$ and $T_{r_2}$), $I_s$ is a null set with probability 1. In
summary, Case (b) cannot contribute anything other than a singleton
to~${\cal A}_0$.

\vskip0.05in\noindent Case (c). Let  $s=s^{(j)}$. The probability
that a specific individual (say, the $i$-th one) has the landmark
event at an age contained in $A_{s^{(j)}}$ is
$P(T_i \in A_{s^{(j)}}, i \in {\cal I}_2)$.
Since this quantity is strictly positive, the probability that none
of the $n$ individuals have had the landmark event in $A_{s^{(j)}}$
and recalled the date is
$\left(1-P(T_i \in A_{s^{(j)}}, i \in {\cal I}_2)\right)^n$,
which goes to zero as $n \rightarrow \infty$. Thus, the probability
that there is $i \in {\cal I}_2$ such that $T_i \in A_{s^{(j)}}$
goes to one as $n\rightarrow \infty$. Therefore, $I_{s^{(j)} \cup
\{i\}}=I_{s^{(j)}} \cap \{T_i\}$ is non-null. Consequently $I_s \in {\cal A}_2$, which means ${\cal A}_0 \subset {\cal A}_2$ almost surely.  It follows that
$P[{\cal A}_2 ={\cal A}_0]$ goes to one as $n\rightarrow \infty$.

The statement of the theorem follows by combining these cases.

\bigskip
{\bf Proof of Theorem 4.4}

From (4.21), the log-likelihood is
\begin{equation}
\ell(\bp,\bE)=\sum_{i=1}^{n}\left(\ln\bigg(\sum_{j=1}^{v}\alpha_{ij}q_j\bigg)\right).
\label{loglik1}
\end{equation}
We maximize $\ell(\bp,\bE)$ periodically with respect to $\bp$ and $\bE$.
If $(\bp^{(n)},\bE^{(n)})$, be the iterate at
the $n$th stage, the next iterate $(\bp^{(n+1)},\bE^{(n+1)})$ is defined by
\begin{eqnarray}
\bE^{(n+1)}&=& \left\{
\begin{array}{ll}
\bE^{(n)} & \text {if $n$ is even},\\
\operatornamewithlimits{argmax}\limits_{\displaystyle\bE\in
M_2}\ell(\bp^{(n)},\bE) & \text {if $n$ is odd},
\end{array} \right.\\
\bp^{(n+1)}&=& \left\{
\begin{array}{ll}
\bp^{(n)} & \text {if $n$ is odd},\\
\operatornamewithlimits{argmax}\limits_{\displaystyle\bp\in
M_1}\ell(\bp,\bE^{(n)}) & \text {if $n$ is even},
\end{array} \right. \label{b,p(n+1)}
\end{eqnarray}
where $M_1=\{\bp:\sum_{j=1}^{v}q_j=1,\ 0\le q_i\le 1, i=1,2,\ldots,v\}$
and $M_2=\{\bE:\eta_i=(b_{i1},b_{i2}, \ldots, b_{ik}),0\leq b_{il}\leq1, \quad \forall i, \forall l\}$. We shall show
that the functions $\ell(\bp,\cdot)$ and $\ell(\cdot,\bE)$ are
concave over the convex sets $M_1$ and $M_2$, respectively, so that
there exists a maximum at each iteration. Thus, in each stage there
is an increase in the likelihood~(4.22), which is bounded
by $(kv)^n$, and the sequence of partially maximized likelihoods
converges. Under the conditions of the theorem, we shall
also show that the objective function is strictly concave, which implies the uniqueness of the
maximum at each stage, with probability tending one when
$n_2$ goes to infinity. Eventually, as $M_1\times M_2$ is a closed set, the sequence of maxima obtained at successive stages converges
to a unique limit, with probability going to one.

Let $\bB$ be an $n\times v$ matrix such that the $ij$th be $\alpha_{ij}$. For fixed $\bb$, the partial derivative of \eqref{loglik1}
with respect to $\bp$ is
$$\frac{\partial \ell}{\partial \bp}=\sum\limits_{i=1}^n\frac{B_i}{{B_i}^T\bp}$$
where $B_i$ is the $i$th row of the matrix $\bB$.
The second derivative or the Hessian is
\begin{equation}
 \frac{\partial \ell}{\partial\bp \partial{\bp}^T}=-\sum\limits_{i=1}^n\frac{ B_iB_i^T}{({B_i}^T\bp)^2} \label{2nd_d_r}
\end{equation}
which is a non-positive definite matrix. Therefore, $\ell$ is a concave function over a convex and bounded domain, which ensures existence of maxima; see \cite{Simon_1994}.

Now, we need to show that the Hessian matrix is negative definite in the long run.
It is enough to show that for any non-zero vector $\bu$,
$$P\left(\sum\limits_{i=1}^n\frac{(B_i^T\bu)^2}{({B_i}^T\bp)^2}=0\right)\rightarrow 0.$$
In other words, we need to show that for any arbitrary non-zero vector $\bu$,
\begin{equation}
P\left( B_i^T \bu=0 \quad   \forall
i\right)=P\left(\bB\bu=0\right)\rightarrow 0.  \label{bu}
\end{equation}

For $i \in {\cal I}_2$, $B_i$ has only one
non-zero element. In this situation, the equation $B_i^T\bu=0$ implies that the corresponding element of $\bu$ is zero.
Moreover, Theorem~4 implies that the intervals  $J_j \in {\cal A}_0$ associated with columns of $\bB$
correspond only to singleton members (${\cal A}_2$) with probability tending to 1. Therefore, with probability tending to one, the event $\bB\bu=0$
coincides with the event $\bu=0$, which has probability zero. 

For fixed $\bp$, the first derivative of \eqref{loglik1} with respect to $\bE$ is
$$\frac{\partial \ell}{\partial \bE}=\sum\limits_{i=1}^n\frac{{\bf A}_i\bp}{{B_i}^T\bp}$$
where ${\bf A}_i$ is the $3k\times m$ matrix with the $(l,j)^{\rm
th}$ element given by $\frac{\partial\alpha_{ij}}{\partial b_{1l}}$, for $l=1,2,\ldots,k$, $\frac{\partial\alpha_{ij}}{\partial b_{2l}}$, for $l=k+1,\ldots,2k$ and $\frac{\partial\alpha_{ij}}{\partial b_{3l}}$, for $l=2k+1,\ldots,3k$.

The Hessian with respect to $\bE$ is
\begin{equation}
\frac{\partial \ell}{\partial\bE
\partial{\bE^T}}=-\sum_{i=1}^n\left(B_i^T\bp\right)^{-2} {\bf
A}_i\bp\bp^T{\bf A}_i^T  \label{2nd_d_d}
\end{equation}
which is non-positive definite matrix. Therefore $\ell$ is a concave
function over a convex domain, which guarantees the existence of a maximum; see \cite{Simon_1994}.

Now, to show the Hessian matrix is negative definite with
probability tending to one, we need to show that for any arbitrary non-zero
vector $\bv$,
\begin{equation}
P\left( \bv^T {\bf A}_i\bp =0 \quad   \forall i\right)\rightarrow 0.
\label{bv}
\end{equation}
For $i\in{\cal I}_2$,
\begin{equation}
 {\bf A}_i \bp=-\left(\sum_{j=1}^v q_j\cdot I(J_j \subset A_i)\right)\big(I(T_i \in A_{i1}),\ldots,I(T_i \in A_{ik})\big)^T, \label{AP}
\end{equation}
which is a vector with a non-zero element exactly at one place. The
condition $\bv^T {\bf A}_i\bp =0$ is equivalent to the requirement
that the element of $\bv$ corresponding to the non-zero element of
${\bf A}_i\bp$ is zero. On the other hand, as $n_2 \rightarrow
\infty$,
\begin{eqnarray*}
&&\hskip-40pt P\left(\sum_{i\in {\cal I}_2} I \big((S_i-T_i) \in
[x_l,x_{l+1}]\big)=0\right)\\
&=&\Big[P\Big((S_i-T_i)\in
[x_l,x_{l+1}]|\delta_i\varepsilon_i=1\Big)\Big]^{n_2} \rightarrow 0
\  \ \  \forall l.
\end{eqnarray*}
Hence, for all $l=1,\ldots,k$, there is at least
one $i\in {\cal I}_2$ such that $T_i\in A_{il}$, with probability
tending to one. Therefore, the condition $\bv^T {\bf A}_i\bp =0
\quad \forall i\in{\cal I}_2$ reduces, with probability tending to
one, to the requirement that all the elements of $\bv$ are zero.
Therefore, for $\bv\ne 0$, we have $P\left( \bv^T {\bf A}_i\bp =0,
\quad   \forall i\right)\le P\left( \bv^T {\bf A}_i\bp =0, \quad
\forall i\in{\cal I}_2\right) \rightarrow 0$. Thus, the probability
that the Hessian matrix defined in~\eqref{2nd_d_d} is negative
definite goes to one. This completes the proof.

\bigskip
{\bf Proof of Theorem 4.5}

 We can incorporate the constraint $\sum\limits_{j=1}^{v} q_j=1$, by using the Lagrange multiplier, to maximize
 \begin{equation}
  \ell=\sum_{i=1}^{n}\left(\ln\bigg(\sum_{j=1}^{v}\alpha_{ij}q_j\bigg)\right)
  +\lambda \left(\sum_{j=1}^{v} q_j -1\right).
 \end{equation}
 By setting the derivative of $\ell$ with respect to $\lambda$ equal to 0, we have
 \begin{equation}
 \frac{\partial \ell}{\partial \lambda}=\sum\limits_{j=1}^{v} q_j-1=0.\\   \label{constrain}
 \end{equation}
 On the other hand, by setting the derivative of $\ell$ with respect to $q_j$s equal to 0, we obtain
 \begin{equation}
  \frac{\partial \ell}{\partial q_j}=\sum\limits_{i=1}^{n}\frac{\alpha_{ij}}{\sum\limits_{r=1}^{v}\alpha_{ir}q_r}-\lambda =0 \qquad\forall j=1,2,\ldots,m.\\  \label{deriv}
 \end{equation}
 By multiplying both sides of \eqref{deriv} by $q_j$ and adding them
 over all values of $j$, we get
 \begin{equation}
  \sum\limits_{j=1}^{v}\sum\limits_{i=1}^{n} \frac{\alpha_{ij}q_j}{\sum\limits_{r=1}^{v}\alpha_{ir}q_r}=\lambda \sum_{j=1}^vq_j,  \label{fir}
 \end{equation}
 which simplifies, after interchange of the summations and utilization of \eqref{constrain}, to
 \begin{equation}
 \lambda=n. \label{lambda}
 \end{equation}
 By substituting into \eqref{deriv} the optimum value of $\lambda$ obtained above, we have
 \begin{equation*}
  \sum_{i=1}^n \frac{\alpha_{ij}}{\sum\limits_{r=1}^v \alpha_{ir}q_r}=n \qquad\mbox{for }j=1,\ldots,v,
 \end{equation*}
that is
 \begin{equation*}
 \frac{1}{n} \sum_{i=1}^n \frac{\alpha_{ij}q_j}{\sum\limits_{r=1}^v \alpha_{ir}q_r}=q_j \qquad\mbox{for }j=1,\ldots,v,
 \end{equation*}
 Thus, $\pi_j(p)=q_j \ \ \forall j$, and the statement is proved.

\bigskip
{\bf Proof of Theorem 4.6}

We use Theorem~3.1 of \cite{Wang_1985}, which was used by \cite{Gentleman_1994}.
There are five assumptions that we need to check in order to establish consistency of the AMLE.

The first assumption requires a separable compactification of the
parameter space $\Theta$. In our case, the set
$\overline{\Theta}$ serves this purpose. For metric we can use the L\'{e}vy distance, and compactness follows from the Helley
selection theorem. In order to establish separability \cite[p.
239]{Billingsley_1968}, we use the Homeomorphic mapping of $[t_{min},t_{max}]$ to
$[0,1]$.

To take care of non-identifiability as in \cite{Redner_1981}, the equivalence class ${\cal E}$ defined by
\begin{equation}
 {\cal E}=\{F\,:\,F\in \Theta,\,E[\ell(F)-\ell(F_0)]=0\},  \label{class}
\end{equation}
is regarded as a single point in $\Theta$.

Let, for $r=1,2,\ldots,$ $V_r(F)$ be the L\'{e}vy neighborhood of
$F\in\Theta$ with radius $1/r$. For such a sequence of decreasing
open neighborhoods, \cite{Wang_1985}'s second assumption requires
that, for any $F_0$ in $\Theta$, there is a function
$F_r:\,\overline{\Theta}\rightarrow V_r(F_0)$ such that (a)
$\ell(F)-\ell(F_r(F))$ is locally dominated on $\overline{\Theta}$
and (b) $F_r(F)$ is in $\Theta$ if $F\in \Theta$. We define
$F_r(F)=\frac{1}{r+1}F+\frac{r}{r+1}F_0$. Since
$\|F_r(F)-F_0\|=\frac{1}{r+1}\|F-F_0\|$, and the L\'{e}vy distance
is dominated by the Kolmogorov-Smirnov distance, it is clear that
$F_r(F)\in V_r(F_0)$. Condition (b) is obviously satisfied. As for
condition (a), note that
\begin{align*}
&\sup_{F\in \overline{\Theta}}\big[\ell(F)-\ell(F_{F,r})\big]\\
&=\sup_{F\in \overline{\Theta}} \ln \frac{\sum_{j=1}^{n_2} \alpha_{ij} \left(F(t_j)-F(t_{j^-})\right)}{\frac{1}{r+1}\big[\sum_{j=1}^{n_2} \alpha_{ij} \left(F(t_j)-F(t_{j^-})\right)\big]+\frac{r}{r+1}\big[\sum_{j=1}^{n_2} \alpha_{ij} \left(F_0(t_j)-F_0(t_{j^-})\right)\big]} \\
& \leq \ln(r+1),
\end{align*}
which has finite expectation. Thus, $\ell(F)-\ell(F_r(F))$ is globally dominated on $\overline{\Theta}$.

The third assumption requires that $E[\ell(F)-\ell(F_r(F))]<0$ for $F_0\in\Theta$, $F\in \overline{\Theta}$, $F\neq F_0$.
Here, $F_0$ needs to be interpreted as ${\cal E}$, and the result follows along the lines of the proof of Lemma 4.4 of \cite{Wang_1985}.

The fourth and fifth assumptions require that $\ell(F)-\ell(F_r(F))$ is lower and upper semicontinuous for $F \in \overline{\Theta}$ except for a null set of points (which may depend on $F$ only in the case of upper semicontinuity). Both the conditions follow from the portmanteau theorem \cite[p. 11]{Billingsley_1968}, as argued by \cite{Gentleman_1994}. No null set needs to be invoked.

The result follows from Theorem 3.1 of \cite{Wang_1985} as all the assumptions hold.

{\bf Proof of Theorem 4.7}

Theorem~4.6 implies that $$\inf_ {F\in{\cal E}}
d_L(\tilde{F_n},F)\rightarrow 0 \qquad \mbox{as } n\rightarrow\infty
\qquad \mbox{with probability 1.} $$ Therefore
$P(\inf_ {F\in{\cal E}} d_L(\tilde{F_n},F) > \epsilon)\rightarrow 0.$

By using Theorem~4.6, we have $P(\omega:
\tilde{F_n}(\omega)=\hat{F_n}(\omega))\rightarrow 1$, and we conclude
$$P\left(\inf_ {F\in{\cal E}} d_L(\hat{F_n},F) > \epsilon\right)\rightarrow 0.$$
\section{Simulation of performance}
{\bf Parametric estimation}:

Tables~\ref{t:tableone} and~\ref{t:tabletwo} show the bias, the standard deviation (Stdev) and the mean squared error (MSE) for the MLEs of the parameter $\theta=(\theta_1,\theta_2)$, the median of time-to-event, and the estimated exact recall probability 5~years after the event, based on the three likelihoods, for the combination of parameter values in case (i) to case (iv) of Section~5.1 of the main paper, for sample size $n=300$ and 1000.
As with simulations with $n=100$ reported in Table~1 of the paper, in cases (i)--(iii), it is found that the bias and the standard deviation (and consequently the MSE) of the Partial Recall MLE is less than those of the other two estimators and its performance improves with increasing sample size. In case (iv) also (the case where the parameters are chosen to produce lesser proportion of partial recalls), it is seen that the overall performance of the proposed Partial Recall MLE is better than that of the Binary Recall MLE, even though for sample size $1000$, the bias of the Binary recall MLE of some parameters is smaller.

\begin{table}
\centering
\caption{Bias, standard deviation (Stdev) and MSE of estimated parameters for $n=300$}
\label{t:tableone}
\scalebox{1}{
\begin{tabular}{ccc@{\hskip3pt}c@{\hskip3pt}cc@{\hskip3pt}c@{\hskip3pt}cc@{\hskip3pt}c@{\hskip3pt}c}
\\[-1.5ex]
\hline
Case	&	Param	&\multicolumn{3}{c}{Current Status MLE}&\multicolumn{3}{c}{Binary Recall MLE}&\multicolumn{3}{c}{Partial Recall MLE}\\
\cline{3-11}
	&		&	Bias	&	Stdev	&	MSE	&	Bias	&	Stdev	&	 MSE	&	Bias	&	Stdev	&	MSE	\\
\hline
(i)	&	$\theta_1$	         &	0.445	&	1.624	&	2.832	&	0.171	&	 0.944	&	0.920	&	0.150	&	0.627	&	0.416	\\
	&	$\theta_2$           &	-0.018	&	0.185	&	0.034	&	-0.005	&	 0.134	&	0.018	&	-0.007	&	0.098	&	0.009	\\
	&	Median	             &	-0.008	&	0.189	&	0.036	&	-0.001	&	 0.020	&	0.020	&	-0.001	&	0.010	&	0.010	\\
	&	$\pi^{(0)}(5)$&	-	&	-	&	-	&	-0.002	&	 0.0301	&	0.0009	&	-0.0006	&	0.030	&	0.0009	\\
\hline
(ii)	&	$\theta_1$       &	0.495	&	1.569	&	2.703	&	0.203	 &	0.898	&	0.847	&	0.128	&	0.533	&	0.300	\\
	&	$\theta_2$	         &	-0.020	&	0.183	&	0.033	&	-0.004	&	 0.130	&	0.017	&	-0.004	&	0.083	&	0.007	\\
	&	Median	             &	-0.007	&	0.188	&	0.035	&	 0.002	&	 0.139	&	0.019	&	-0.0002	&	0.087	&	0.007	\\
	&	$\pi^{(0)}(5)$&	-	    &	-	    &	-	    &	-0.013	&	 0.032	&	0.0002	&	0.002	&	0.034	&	0.001	\\
\hline
(iii)	&	$\theta_1$	     &	0.149	&	1.519	&	2.504	&	0.149	&	0.954	&	0.932	&	0.184	&	0.720	&	0.553	\\
	&	$\theta_2$	         &	-0.023	&	0.186	&	0.035	&	-0.009	&	 0.138	&	0.020	&	0.001	&	0.104	&	0.011	\\
	&	Median	             &	-0.012	&	0.189	&	0.036	&	-0.006	&	 0.149	&	0.022	&	0.006	&	0.113	&	0.013	\\
	&	$\pi^{(0)}(5)$&	-	    &	-	    &	-	    &	-0.020	&	 0.031	&	0.001	&	0.001	&	0.035	&	0.001	\\
\hline
(iv)	&	$\theta_1$	     &	0.583	&	1.660	&	3.092	&	0.157	 &	0.752	&	0.591	&	0.137	&	0.667	&	0.464	\\
	&	$\theta_2$	         &	-0.019	&	0.187   &	0.035   &	-0.002	&	 0.108	&	0.012	&	-0.005	&	0.096	&	0.009	\\
	&	Median	             &	-0.004	&	0.191	&	0.036	&	 0.002	&	 0.117	&	0.013	&	-0.0008	&	0.104	&	0.011	\\
	&	$\pi^{(0)}(5)$&	-	    &	-	    &	-	    &	-0.012	&	 0.037	&	0.001	&	0.003	&	0.038	&	0.001	\\
\hline
\end{tabular}
}
\end{table}

\begin{table}
\centering
\caption{Bias, standard deviation (Stdev) and MSE of estimated parameters for $n=1000$}
\label{t:tabletwo}
\scalebox{1}{
\begin{tabular}{ccc@{\hskip3pt}c@{\hskip3pt}cc@{\hskip3pt}c@{\hskip3pt}cc@{\hskip3pt}c@{\hskip3pt}c}
\\[-1.5ex]
\hline
Case	&	Param	&\multicolumn{3}{c}{Current Status MLE}&\multicolumn{3}{c}{Binary Recall MLE}&\multicolumn{3}{c}{Partial Recall MLE}\\
\cline{3-11}
	&		&	Bias	&	Stdev	&	MSE	&	Bias	&	Stdev	&	 MSE	&	Bias	&	Stdev	&	MSE	\\
\hline
(i)	&	$\theta_1$	&	0.119	&	0.811	&	0.672	&	0.037	&	 0.497	&	0.248	&	0.09	&	0.329	&	0.116	\\
	&	$\theta_2$	&	-0.007	&	0.103	&	0.011	&	-0.004	&	 0.075	&	0.006	&	-0.003	&	0.053	&	0.002	\\
	&	Median	&	-0.005	&	0.106	&	0.011	&	-0.003	&	 0.08	&	0.006	&	-0.0001	&	0.056	&	0.003	\\
	&	$\pi_{\eta}^{(0)}(5)$	&	-	&	-	&	-	&	-0.0002	&	 0.0161	&	0.0002	&	0.0003	&	0.016	&	0.0002	\\
\hline
(ii)	&	$\theta_1$	&	0.123	&	0.803	&	0.66	&	0.046	 &	0.483	&	0.235	&	0.038	&	0.288	&	0.09	\\
	&	$\theta_2$	&	-0.007	&	0.103	&	0.011	&	-0.002	&	 0.071	&	0.005	&	-0.002	&	0.046	&	0.002	\\
	&	Median	&	-0.004	&	0.105	&	0.011	&	-0.001	&	 0.077	&	0.006	&	-0.0008	&	0.048	&	0.002	\\
	&	$\pi_{\eta}^{(0)}(5)$	&	-	&	-	&	-	&	-0.014	&	 0.017	&	0.0005	&	0.0004	&	0.017	&	0.0003	\\
\hline
(iii)	&	$\theta_1$	&	0.123	&	0.842	&	0.723	&	0.043	 &	0.528	&	0.281	&	0.107	&	0.395	&	0.167	\\
	&	$\theta_2$	&	-0.007	&	0.103	&	0.011	&	-0.002	&	 0.077	&	0.006	&	-0.002	&	0.057	&	0.003	\\
	&	Median	&	-0.005	&	0.105	&	0.011	&	-0.001	&	 0.083	&	0.007	&	0.001	&	0.062	&	0.003	\\
	&	$\pi_{\eta}^{(0)}(5)$	&	-	&	-	&	-	&	-0.019	&	 0.017	&	0.0006	&	-0.0002	&	0.017	&	0.0003	\\
\hline
(iv)	&	$\theta_1$	&	0.108	&	0.798	&	0.649	&	0.036	 &	0.406	&	0.166	&	0.069	&	0.358	&	0.133	\\
	&	$\theta_2$	&	-0.009	&	0.101	&	0.01	&	-0.0001	&	 0.06	&	0.003	&	-0.0007	&	0.05	&	0.002	\\
	&	Median	&	-0.006	&	0.103	&	0.011	&	0.001	&	 0.064	&	0.004	&	0.001	&	0.057	&	0.003	\\
	&	$\pi_{\eta}^{(0)}(5)$	&	-	&	-	&	-	&	-0.012	&	 0.02	&	0.0005	&	0.002	&	0.02	&	0.0004	\\
\hline
\end{tabular}
}
\end{table}

For sensitivity analysis, a mixture model of Weibull and Log-Normal is considered for the time-to-event distribution. The simulation set-up is described in Section~5.1 of the paper. Tables~\ref{t:tablethree} and~\ref{t:tablefour} show for $\gamma=0.2$ and $\gamma=0.5$, respectively, the bias, the standard deviation (Stdev) and the mean squared error (MSE) for the MLEs of the parameter $\theta=(\theta_1,\theta_2)$, the median of time-to-event, and the estimated exact recall probability 5~years after the event, based on the three likelihoods, for the mixture model for time-to-event introduced in last paragraph of Section~5.1 and the combination of parameter values in case (i) to case (iv) of Section~5.1 of the main paper, for sample size $n=300$. The choice of sample size is in line with the real data.
This miss-specification increases the bias in Partial recall MLE and Binary Recall MLE more than the same in Current Status MLE especially when $\gamma=0.5$, though its standard deviation remains smaller. Partial Recall MLE has the smallest MSE when  $\gamma=0.2$. For $\gamma=0.5$, both Partial Recall MLE and Binary Recall MLE have larger MSE than the Current Status MLE in the cases (iii) and (iv).
\begin{table}
\centering
\caption{Bias, standard deviation (Stdev) and MSE of estimated parameters for $n=300, \gamma=0.2$,}
\label{t:tablethree}
\scalebox{0.9}{
\begin{tabular}{ccc@{\hskip3pt}c@{\hskip3pt}cc@{\hskip3pt}c@{\hskip3pt}cc@{\hskip3pt}c@{\hskip3pt}c}
\\[-1.5ex]
\hline
Case	&	Param	&\multicolumn{3}{c}{Current Status MLE}&\multicolumn{3}{c}{Binary Recall MLE}&\multicolumn{3}{c}{Partial Recall MLE}\\
\cline{3-11}
	&		&	Bias	&	Stdev	&	MSE	&	Bias	&	Stdev	&	 MSE	&	Bias	&	Stdev	&	MSE	\\
\hline
(i)	    &	$\theta_1$	&	0.985	&	1.851	&	4.392	&	0.553	 &	1.033   &	1.372	&	0.533	&	0.714	&	0.794\\
	    &	$\theta_2$	&	0.097	&	0.192	&	0.046	&	0.104	 &	0.138	&	0.030	&	0.092&	0.099	&	0.018	\\
	    &	Median	    &	0.121	&	0.184	&	0.049	&	0.119	&	 0.135	&	0.032	&	0.109	&	0.096	&	0.021	\\
	    &	$\pi_{\eta}^{(0)}(5)$	&	-	&	-	&	-	&	-0.001	 &	0.030	&	0.001	&	0.0003	&	0.031	&	0.001	\\
\hline
(ii)	&	$\theta_1$	&	0.991	&	1.961	&	4.824	&	0.416	 &	0.990	&	1.152   &	0.502	&	0.614	&	0.628	\\
	    &	$\theta_2$	&	0.089	&	0.193	&	0.045	&	0.089    &	0.133	&	0.025	&	0.094	&	0.087	&	0.016	\\
	    &	Median   	&	0.113	&	0.189	&	0.048	&	0.100	 &	 0.134	&	0.027	&	0.109	&	0.084	&	0.019	\\
	    &	$\pi_{\eta}^{(0)}(5)$	&	-	&	-	&	-	&	-0.014	 &	0.031	&	0.001	&	0.0003	&	0.033	&	0.001	\\
\hline
(iii)	&	$\theta_1$	&	0.953	&	1.971	&	4.791	&	0.407	 &	1.113	&	1.403	&	0.403	&	0.838	&	0.863	\\
	    &	$\theta_2$	&	0.101	&	0.189	&	0.046	&	0.095	 &	0.135	&	0.027	&	0.089	&	0.100	&	0.018	\\
	    &	Median	    &	0.123	&	0.184	&	0.048	&	0.103   &	 0.139	&	0.030	&	0.099	&	0.103	&	0.020	\\
	    &	$\pi_{\eta}^{(0)}(5)$	&	-	&	-	&	-	&	-0.020	 &	0.031	&	0.001	&	0.001	&	0.035	&	0.001	\\
\hline
(iv)	&	$\theta_1$	&	1.081	&	1.888	&	4.729	&	0.378	 &	0.844	&	0.854	&	0.383	&	0.746	&	0.702	\\
	    &	$\theta_2$	&	0.082	&	0.190	&	0.043	&	0.086	 &	0.110	&	0.019	&	0.085	&	0.099	&	0.017	\\
	    &	Median	    &	0.110	&	0.185	&	0.046	&	0.096	&	 0.111 &	0.022	&	0.095	&	0.100	&	0.019	\\
	    &	$\pi_{\eta}^{(0)}(5)$	&	-	&	-	&	-	&	-0.010	 &	0.036	&	0.001	&	0.004	&	0.037	&	0.001	\\
\hline
\end{tabular}
}
\end{table}

\begin{table}
\centering
\caption{Bias, standard deviation (Stdev) and MSE of estimated parameters for $n=300, \gamma=0.5$}
\label{t:tablefour}
\scalebox{0.9}{
\begin{tabular}{ccc@{\hskip3pt}c@{\hskip3pt}cc@{\hskip3pt}c@{\hskip3pt}cc@{\hskip3pt}c@{\hskip3pt}c}
\\[-1.5ex]
\hline
Case	&	Param	&\multicolumn{3}{c}{Current Status MLE}&\multicolumn{3}{c}{Binary Recall MLE}&\multicolumn{3}{c}{Partial Recall MLE}\\
\cline{3-11}
	&		&	Bias	&	Stdev	&	MSE	&	Bias	&	Stdev	&	 MSE	&	Bias	&	Stdev	&	MSE	\\
\hline
(i)	    &	$\theta_1$	&	-2.812	&	1.176	&	9.296	&	-1.934	 &	0.226	&	3.794	&	-1.967	&	0.129	&	3.886	\\
	    &	$\theta_2$	&	0.516	&	0.245   &	0.327	&	0.561	 &	0.228   &	0.367	&	0.515	&	0.178	&	0.297	\\
	    &	Median	    &	0.311	&	0.226	&	0.148	&	0.434	&	 0.215	&	0.235	&	0.389	&	0.168	&	0.179	\\
	    &	$\pi_{\eta}^{(0)}(5)$	&	-	&	-	&	-	&	-0.00002 &	0.031	&	0.001	&	0.001	&	0.031	&	0.001	\\
\hline
(ii)	&	$\theta_1$	&	-2.863	&	1.098	&	9.403	&	-1.963	 &	0.152	&	3.877	&	-1.979	&	0.093	&	3.927	\\
	    &	$\theta_2$	&	0.505	&	0.236	&	0.311	&	0.0574	 &	0.228	&	0.382	&	0.506	&	0.159	&	0.281	\\
	    &	Median  	&	0.297	&	0.221	&	0.137	&	0.445	&	 0.215 &	0.245	&	0.379	&	0.150	&	0.167	\\
	    &	$\pi_{\eta}^{(0)}(5)$	&	-	&	-	&	-	&	-0.017	 &	0.034	&	0.001	&	-0.0005	&	0.035	&	0.001	\\
\hline
(iii)	&	$\theta_1$	&	-2.784	&	1.207	&	9.209	&	-1.963	 &	0.171	&	3.886	&	-1.988	&	0.073	&	3.959	\\
	    &	$\theta_2$	&	0.507	&	0.247	&	0.317	&	0.583	 &	0.228	&	0.392	&	0.547	&	0.184	&	0.333	\\
	    &	Median  	&	0.303	&	0.226	&	0.143	&	0.454	&	 0.215	&	0.252	&	0.418	&	0.175	&	0.205	\\
	    &	$\pi_{\eta}^{(0)}(5)$	&	-	&	-	&	-	&	-0.022	 &	0.032	&	0.001	&	-0.0002	&	0.036	&	0.001	\\
\hline
(iv)	&	$\theta_1$	&	-2.760	&	1.197	&	9.050	&	-1.978	 &	0.113	&	3.926	&	-1.986	&	0.083	&	3.950	\\
	    &	$\theta_2$	&	0.505	&	0.246	&	0.315	&	0.560    &	0.197	&	0.352	&	0.543	&	0.176	&	0.325	\\
	    &	Median	    &	0.304	&	0.228	&	0.147	&	0.431	&	 0.187	&	0.220	&	0.414	&	0.167	&	0.199	\\
	    &	$\pi_{\eta}^{(0)}(5)$	&	-	&	-	&	-	&	-0.015	 &	0.037	&	0.002	&	-0.001	&	0.038	&	0.001
	\\
\hline
\end{tabular}
}
\end{table}

\begin{figure}[tbh]
\centering
\includegraphics[width=\textwidth]{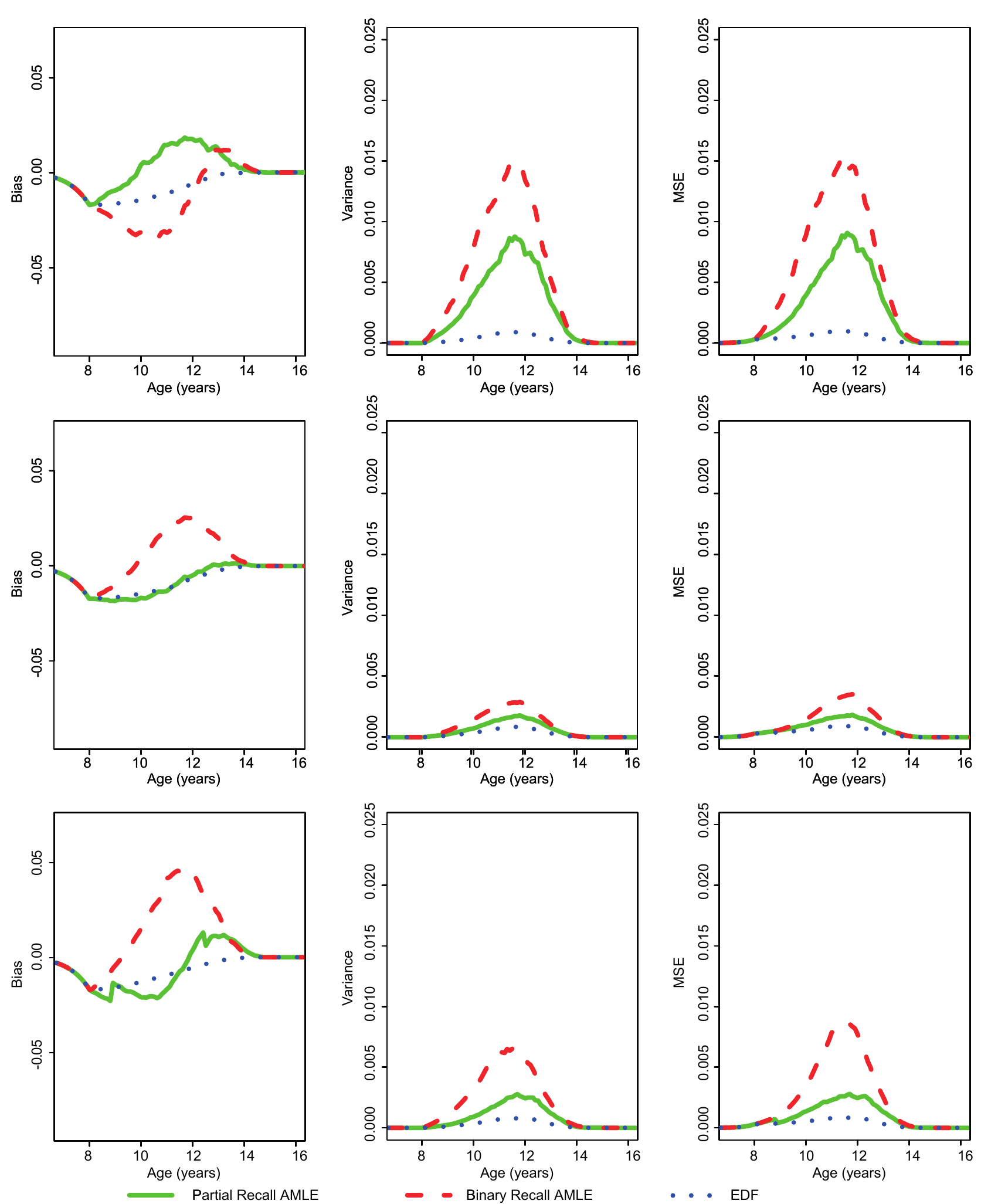}
\caption[]{Comparison of bias, variance and MSE of the estimator for $n=300$ in cases (a) (top panel), (b) (middle panel) and (c) (bottom panel)} \label{fig_300}
\end{figure}

\begin{figure}[tbh]
\centering
\includegraphics [width=\textwidth]{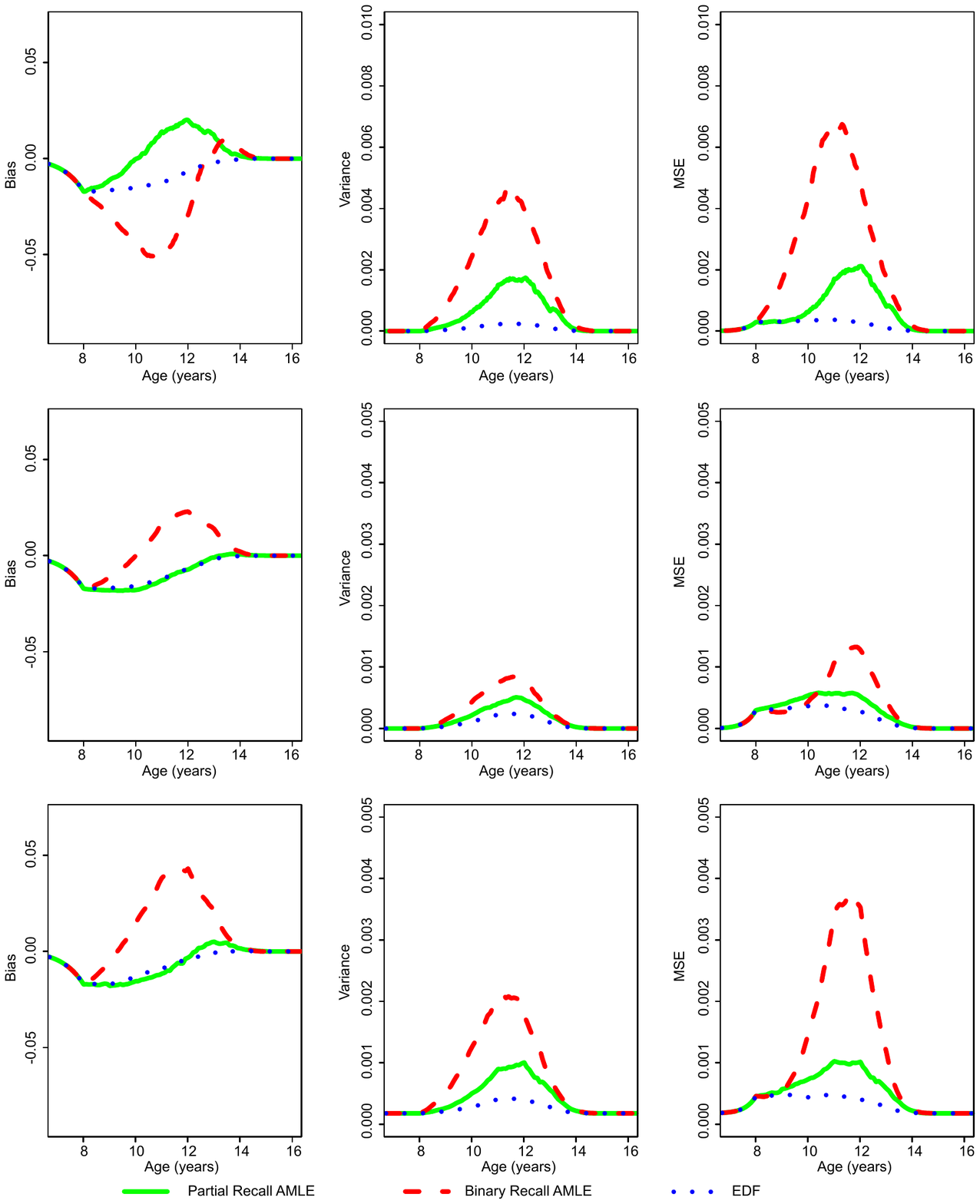}
\caption[]{Comparison of bias, variance and MSE of the estimator for $n=1000$ in cases (a) (top panel), (b) (middle panel) and (c) (bottom panel)} \label{fig_1000}
\end{figure}

\bigskip
{\bf Non-Parametric estimation}:

Figures~\ref{fig_300} and~\ref{fig_1000} show the plots of the bias, the variance and the mean square error (MSE) of the three estimators for different ages, when $n=300$ and 1000 and the parameters of the recall probability functions (4.12) are chosen as in Cases (a), (b) and (c).
The Partial Recall AMLE is found to have smaller bias, variance and MSE than the Binary Recall AMLE estimator, although its performance is expectedly poorer than that of EDF. In contrast with the case of $n=100$ reported in the paper, the performance parameters of the Partial Recall AMLE are found to be closer to those of the EDF (the benchmark usable only for complete data) than with those of the Binary Recall AMLE.




\section{Data analysis}
\begin{figure}
\centering
 \includegraphics[height=4in,width=2.5in,angle=-90]{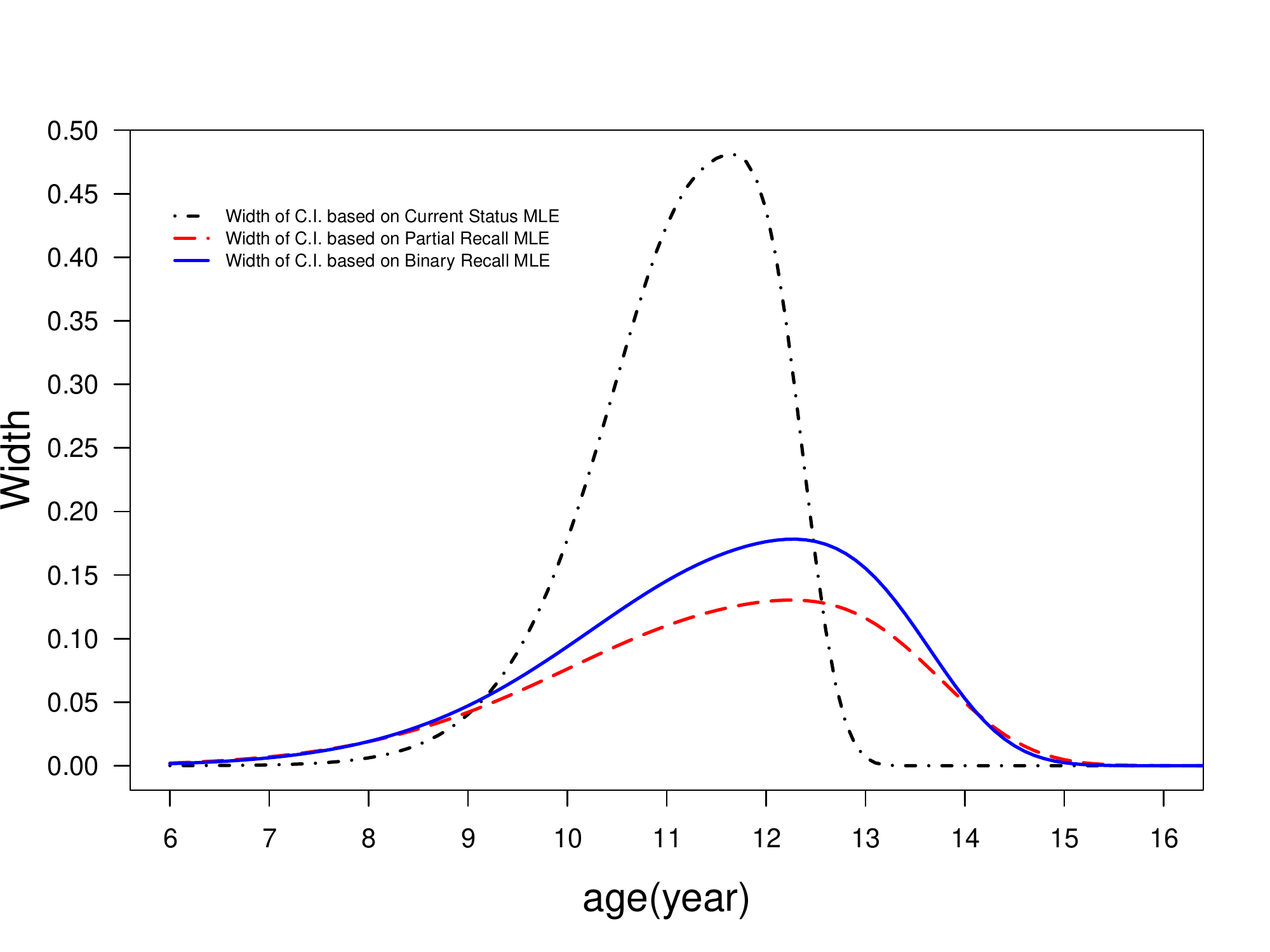}
\caption[]{Width of asymptotic pointwise $95\%$ confidence intervals of survival function for the menarcheal data based on three parametric methods} \label{fig_3}
\end{figure}

For comparison of performance of the proposed Partial Recall MLE with Binary Recall MLE and Current Status MLE, Figure~\ref{fig_3} shows the plots of the widths of the asymptotic pointwise $95\%$ confidence intervals of the estimated survival function based on the three likelihoods. It is clear that the confidence intervals for the Partial Recall MLE are narrower.

\begin{figure}[tbh]
\centering
 \includegraphics[height=3.6in,width=2.8in,angle=-90]{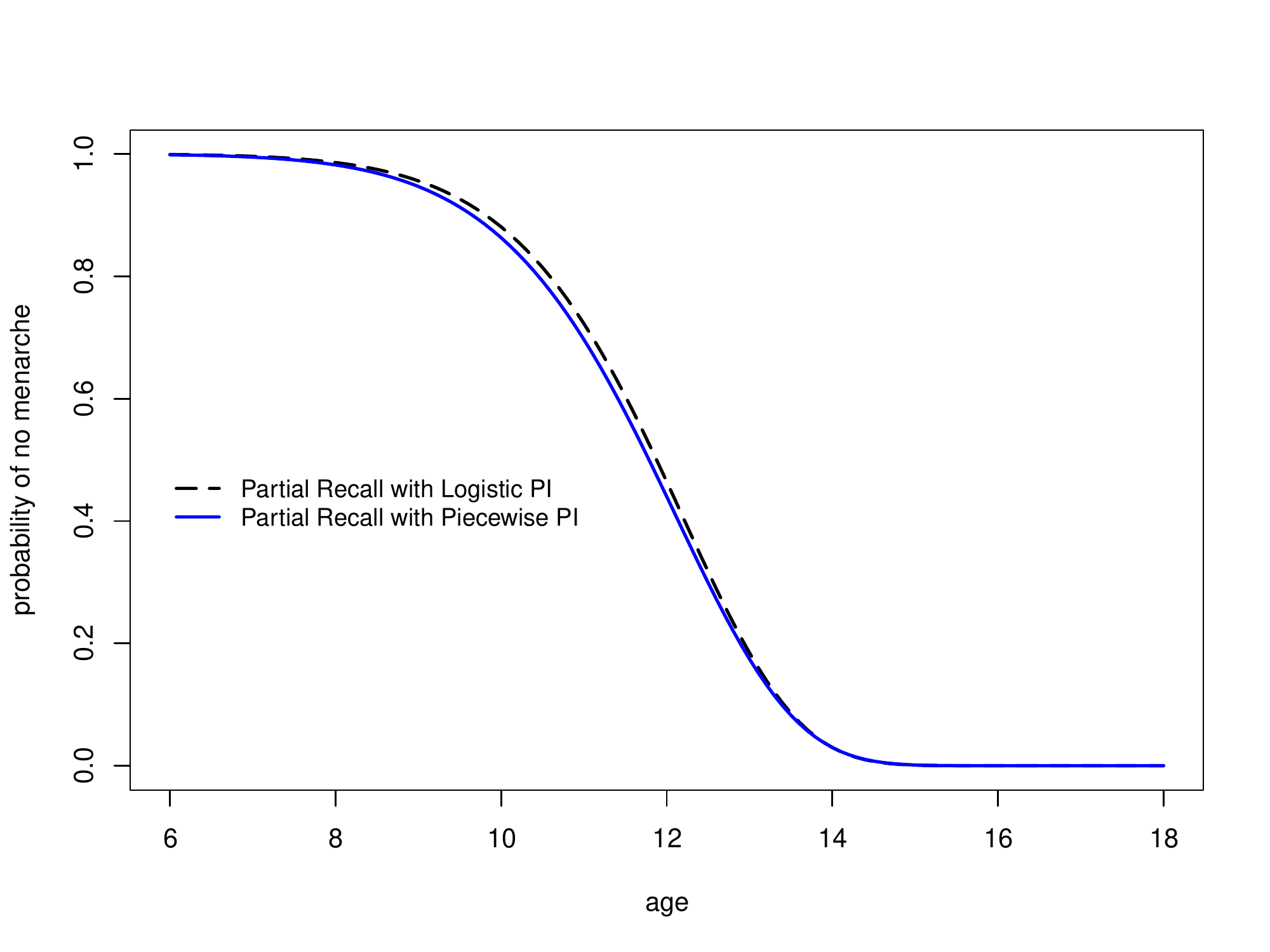}
\caption[]{Survival functions for the menarcheal data based on two different model for recall probability} \label{fig_4}
\end{figure}

To check the adequacy of the functional form of the recall probability, the survival functions of time to menarche is estimated by utilizing the proposed method and considering both multiple logistic regression model presented in Section~5.1 and the piece-wise constant recall probability model introduced in~4.12 of main paper (with knot points of the recall probability functions chosen as in the first paragraph of Section~5.2. Figure~\ref{fig_4} shows the survival functions plot of age at menarche that are very close to each other.  

\newpage


\bibliographystyle{biorefs}
\bibliography{refs}